\documentclass[twocolumn]{aastex631}
\pdfoutput=1
\usepackage{lineno}
\submitjournal{The Astronomical Journal}

\received{December 2, 2022}
\revised{July 5, 2023}
\accepted{July 18, 2023}

\shorttitle{OGLE-2019-BLG-0825}
\shortauthors{Satoh et al. (2023)}

\begin{document}
\title{OGLE-2019-BLG-0825: Constraints on the Source System and Effect on Binary-lens Parameters arising from a Five Day Xallarap Effect in a Candidate Planetary Microlensing Event}

\author[0000-0002-1228-4122]{Yuki K. Satoh}
\correspondingauthor{Yuki K. Satoh}
\email{satohyk@iral.ess.sci.osaka-u.ac.jp}
\altaffiliation{The MOA Collaboration}
\affiliation{Department of Earth and Space Science, Graduate School of Science, Osaka University, 1-1 Machikaneyama-cho, Toyonaka, Osaka 560-0043, Japan}
\affiliation{College of Science and Engineering, Kanto Gakuin University, 1-50-1 Mutsuurahigashi, Kanazawa-ku, Yokohama, Kanagawa 236-8501, Japan}

\author[0000-0003-2302-9562]{Naoki Koshimoto}
\altaffiliation{The MOA Collaboration}
\affiliation{Department of Earth and Space Science, Graduate School of Science, Osaka University, 1-1 Machikaneyama-cho, Toyonaka, Osaka 560-0043, Japan}

\author[0000-0001-8043-8413]{David P. Bennett}
\altaffiliation{The MOA Collaboration}
\affiliation{Code 667, NASA Goddard Space Flight Center, Greenbelt, MD 20771, USA}
\affiliation{Department of Astronomy, University of Maryland, College Park, MD 20742, USA}

\author[0000-0002-4035-5012]{Takahiro Sumi}
\altaffiliation{The MOA Collaboration}
\affiliation{Department of Earth and Space Science, Graduate School of Science, Osaka University, 1-1 Machikaneyama-cho, Toyonaka, Osaka 560-0043, Japan}

\author[0000-0001-5069-319X]{Nicholas J. Rattenbury}
\altaffiliation{The MOA Collaboration}
\affiliation{Department of Physics, University of Auckland, Private Bag 92019, Auckland, New Zealand}

\author[0000-0002-5843-9433]{Daisuke Suzuki}
\altaffiliation{The MOA Collaboration}
\affiliation{Department of Earth and Space Science, Graduate School of Science, Osaka University, 1-1 Machikaneyama-cho, Toyonaka, Osaka 560-0043, Japan}

\author[0000-0001-9818-1513]{Shota Miyazaki}
\altaffiliation{The MOA Collaboration}
\affiliation{Institute of Space and Astronautical Science, Japan Aerospace Exploration Agency, Kanagawa 252-5210, Japan}

\author{Ian A. Bond}
\altaffiliation{The MOA Collaboration}
\affiliation{Institute of Natural and Mathematical Sciences, Massey University, Auckland 0745, New Zealand}

\author[0000-0001-5207-5619]{Andrzej Udalski}
\altaffiliation{The OGLE Collaboration}
\affiliation{Astronomical Observatory, University of Warsaw, Al. Ujazdowskie 4, 00-478 Warszawa, Poland}

\author{Andrew Gould}
\altaffiliation{The KMTNet Collaboration}
\affiliation{Max Planck Institute for Astronomy, K\"{o}nigstuhl 17, D-69117, Heidelberg, Germany}
\affiliation{Department of Astronomy, Ohio State University, 140 W. 18th Avenue, Columbus, OH 43210, USA}

\author[0000-0003-4590-0136]{Valerio Bozza}
\altaffiliation{The MiNDSTEp Collaboration}
\affiliation{Dipartimento di Fisica “E.R. Canianiello,” Università di Salerno, Via Giovanni Paolo II 132, I-84084 Fisciano, Italy}
\affiliation{Istituto Nazionale di Fisica Nucleare, Sezione di Napoli, Napoli, Italy}

\author[0000-0002-3202-0343]{Martin Dominik}
\altaffiliation{The MiNDSTEp Collaboration}
\affiliation{University of St Andrews, Centre for Exoplanet Science, SUPA School of Physics \& Astronomy, North Haugh, St Andrews, KY16 9SS, UK}

\author[0000-0003-4776-8618]{Yuki Hirao}
\altaffiliation{The MOA Collaboration}
\affiliation{Institute of Astronomy, Graduate School of Science, The University of Tokyo, 2-21-1 Osawa, Mitaka, Tokyo 181-0015, Japan}

\author[0000-0002-3401-1029]{Iona Kondo}
\altaffiliation{The MOA Collaboration}
\affiliation{Department of Earth and Space Science, Graduate School of Science, Osaka University, 1-1 Machikaneyama-cho, Toyonaka, Osaka 560-0043, Japan}

\author{Rintaro Kirikawa}
\altaffiliation{The MOA Collaboration}
\affiliation{Department of Earth and Space Science, Graduate School of Science, Osaka University, 1-1 Machikaneyama-cho, Toyonaka, Osaka 560-0043, Japan}

\author{Ryusei Hamada}
\affiliation{Department of Earth and Space Science, Graduate School of Science, Osaka University, 1-1 Machikaneyama-cho, Toyonaka, Osaka 560-0043, Japan}

\collaboration{16}{(Leading Authors)}

\author{Fumio Abe}
\affiliation{Institute for Space-Earth Environmental Research, Nagoya University, Nagoya 464-8601, Japan}

%
\author[0000-0003-4916-0892]{Richard Barry}
\affiliation{Code 667, NASA Goddard Space Flight Center, Greenbelt, MD 20771, USA}

\author{Aparna Bhattacharya}
\affiliation{Code 667, NASA Goddard Space Flight Center, Greenbelt, MD 20771, USA}
\affiliation{Department of Astronomy, University of Maryland, College Park, MD 20742, USA}

\author{Hirosane Fujii}
\affiliation{Department of Earth and Space Science, Graduate School of Science, Osaka University, 1-1 Machikaneyama-cho, Toyonaka, Osaka 560-0043, Japan}

\author[0000-0002-4909-5763]{Akihiko Fukui}
\affiliation{Department of Earth and Planetary Science, Graduate School of Science, The University of Tokyo, 7-3-1 Hongo, Bunkyo-ku, Tokyo 113-0033, Japan}
\affiliation{Instituto de Astrofísica de Canarias, Vía Láctea s/n, E-38205 La Laguna, Tenerife, Spain}

\author{Katsuki Fujita}
\affiliation{Department of Earth and Space Science, Graduate School of Science, Osaka University, 1-1 Machikaneyama-cho, Toyonaka, Osaka 560-0043, Japan}
%

%
%
\author{Tomoya Ikeno}
\affiliation{Department of Earth and Space Science, Graduate School of Science, Osaka University, 1-1 Machikaneyama-cho, Toyonaka, Osaka 560-0043, Japan}

\author[0000-0003-2267-1246]{Stela {Ishitani~Silva}}
\affiliation{Code 667, NASA Goddard Space Flight Center, Greenbelt, MD 20771, USA}
\affiliation{Department of Physics, The Catholic University of America, Washington, DC 20064, USA}

\author[0000-0002-8198-1968]{Yoshitaka Itow}
\affiliation{Institute for Space-Earth Environmental Research, Nagoya University, Nagoya 464-8601, Japan}

%
\author{Yutaka Matsubara}
\affiliation{Institute for Space-Earth Environmental Research, Nagoya University, Nagoya 464-8601, Japan}

\author{Sho Matsumoto}
\affiliation{Department of Earth and Space Science, Graduate School of Science, Osaka University, 1-1 Machikaneyama-cho, Toyonaka, Osaka 560-0043, Japan}

\author[0000-0003-1978-2092]{Yasushi Muraki}
\affiliation{Institute for Space-Earth Environmental Research, Nagoya University, Nagoya 464-8601, Japan}

\author{Kosuke Niwa}
\affiliation{Department of Earth and Space Science, Graduate School of Science, Osaka University, 1-1 Machikaneyama-cho, Toyonaka, Osaka 560-0043, Japan}

\author{Arisa Okamura}
\affiliation{Department of Earth and Space Science, Graduate School of Science, Osaka University, 1-1 Machikaneyama-cho, Toyonaka, Osaka 560-0043, Japan}

\author[0000-0001-8472-2219]{Greg Olmschenk}
\affiliation{Code 667, NASA Goddard Space Flight Center, Greenbelt, MD 20771, USA}

\author[0000-0003-2388-4534]{Cl$\acute{\rm e}$ment Ranc}
\affiliation{Sorbonne Universit\'e, CNRS, UMR 7095, Institut d'Astrophysique de Paris, 98 bis bd Arago, 75014 Paris, France}

\author{Taiga Toda}
\affiliation{Department of Earth and Space Science, Graduate School of Science, Osaka University, 1-1 Machikaneyama-cho, Toyonaka, Osaka 560-0043, Japan}

\author{Mio Tomoyoshi}
\affiliation{Department of Earth and Space Science, Graduate School of Science, Osaka University, 1-1 Machikaneyama-cho, Toyonaka, Osaka 560-0043, Japan}

\author{Paul J. Tristram}
\affiliation{University of Canterbury Mt. John Observatory, P.O. Box 56, Lake Tekapo 8770, New Zealand}

\author[0000-0002-9881-4760]{Aikaterini Vandorou}
\affiliation{Code 667, NASA Goddard Space Flight Center, Greenbelt, MD 20771, USA}
\affiliation{Department of Astronomy, University of Maryland, College Park, MD 20742, USA}

\author{Hibiki Yama}
\affiliation{Department of Earth and Space Science, Graduate School of Science, Osaka University, 1-1 Machikaneyama-cho, Toyonaka, Osaka 560-0043, Japan}

\author{Kansuke Yamashita}
\affiliation{Department of Earth and Space Science, Graduate School of Science, Osaka University, 1-1 Machikaneyama-cho, Toyonaka, Osaka 560-0043, Japan}

%
\collaboration{22}{(The MOA Collaboration)}

\author[0000-0001-7016-1692]{Przemek Mr$\acute{\rm o}$z}
\affiliation{Astronomical Observatory, University of Warsaw, Al. Ujazdowskie 4, 00-478 Warszawa, Poland}

\author[0000-0002-9245-6368]{Radosław Poleski}
\affiliation{Astronomical Observatory, University of Warsaw, Al. Ujazdowskie 4, 00-478 Warszawa, Poland}

\author[0000-0002-2335-1730]{Jan Skowron}
\affiliation{Astronomical Observatory, University of Warsaw, Al. Ujazdowskie 4, 00-478 Warszawa, Poland}

\author[0000-0002-0548-8995]{Michał K. Szyma$\acute{\rm n}$ski}
\affiliation{Astronomical Observatory, University of Warsaw, Al. Ujazdowskie 4, 00-478 Warszawa, Poland}

\author{Radek Poleski}
\affiliation{Astronomical Observatory, University of Warsaw, Al. Ujazdowskie 4, 00-478 Warszawa, Poland}

\author[0000-0002-7777-0842]{Igor Soszy$\acute{\rm n}$ski}
\affiliation{Astronomical Observatory, University of Warsaw, Al. Ujazdowskie 4, 00-478 Warszawa, Poland}

\author[0000-0002-2339-5899]{Paweł Pietrukowicz}
\affiliation{Astronomical Observatory, University of Warsaw, Al. Ujazdowskie 4, 00-478 Warszawa, Poland}

\author[0000-0003-4084-880X]{Szymon Kozłowski}
\affiliation{Astronomical Observatory, University of Warsaw, Al. Ujazdowskie 4, 00-478 Warszawa, Poland}

\author[0000-0001-6364-408X]{Krzysztof Ulaczyk}
\affiliation{Department of Physics, University of Warwick, Gibbet Hill Road, Coventry, CV4 7AL, UK}

\author[0000-0002-9326-9329]{Krzysztof A. Rybicki}
\affiliation{Astronomical Observatory, University of Warsaw, Al. Ujazdowskie 4, 00-478 Warszawa, Poland}
\affiliation{Department of Particle Physics and Astrophysics, Weizmann Institute of Science, Rehovot 76100, Israel}

\author[0000-0002-6212-7221]{Patryk Iwanek}
\affiliation{Astronomical Observatory, University of Warsaw, Al. Ujazdowskie 4, 00-478 Warszawa, Poland}

\author[00000-0002-3051-274X]{Marcin Wrona}
\affiliation{Astronomical Observatory, University of Warsaw, Al. Ujazdowskie 4, 00-478 Warszawa, Poland}

\author[0000-0002-1650-1518]{Mariusz Gromadzki}
\affiliation{Astronomical Observatory, University of Warsaw, Al. Ujazdowskie 4, 00-478 Warszawa, Poland}

\collaboration{13}{(The OGLE Collaboration)}

\author[0000-0003-3316-4012]{Michael D. Albrow}
\affiliation{University of Canterbury, Department of Physics and Astronomy, Private Bag 4800, Christchurch 8020, New Zealand}

\author[0000-0001-6285-4528]{Sun-Ju Chung}
\affiliation{Korea Astronomy and Space Science Institute, Daejon 34055, Republic of Korea}

\author[0000-0002-2641-9964]{Cheongho Han}
\affiliation{Department of Physics, Chungbuk National University, Cheongju 28644, Republic of Korea}

\author[0000-0002-9241-4117]{Kyu-Ha Hwang}
\affiliation{Korea Astronomy and Space Science Institute, Daejon 34055, Republic of Korea}

\author{Doeon Kim}
\affiliation{Department of Physics, Chungbuk National University, Cheongju 28644, Republic of Korea}

\author{Youn Kil Jung}
\affiliation{Korea Astronomy and Space Science Institute, Daejon 34055, Republic of Korea}
\affiliation{University of Science and Technology, Korea, (UST), 217 Gajeong-ro, Yuseong-gu, Daejeon 34113, Republic of Korea}

\author{Hyoun Woo Kim}
\affiliation{Korea Astronomy and Space Science Institute, Daejon 34055, Republic of Korea}

\author[0000-0001-9823-2907]{Yoon-Hyun Ryu}
\affiliation{Korea Astronomy and Space Science Institute, Daejon 34055, Republic of Korea}

\author[0000-0002-4355-9838]{In-Gu Shin}
\affiliation{Center for Astrophysics $|$ Harvard $\&$ Smithsonian, 60 Garden Street,Cambridge, MA 02138, USA}

\author[0000-0003-1525-5041]{Yossi Shvartzvald}
\affiliation{Department of Particle Physics and Astrophysics, Weizmann Institute of Science, Rehovot 76100, Israel}

\author[0000-0003-0626-8465]{Hongjing Yang}
\affiliation{Department of Astronomy and Tsinghua Centre for Astrophysics, Tsinghua University, Beijing 100084, People’s Republic of China}

\author[0000-0001-9481-7123]{Jennifer C. Yee}
\affiliation{Center for Astrophysics $|$ Harvard $\&$ Smithsonian, 60 Garden Street,Cambridge, MA 02138, USA}

\author[0000-0001-6000-3463]{Weicheng Zang}
\affiliation{Department of Astronomy and Tsinghua Centre for Astrophysics, Tsinghua University, Beijing 100084, People’s Republic of China}

\author{Sang-Mok Cha}
\affiliation{Korea Astronomy and Space Science Institute, Daejon 34055, Republic of Korea}
\affiliation{School of Space Research, Kyung Hee University, Yongin, Kyeonggi 17104, Republic of Korea}

\author{Dong-Jin Kim}
\affiliation{Korea Astronomy and Space Science Institute, Daejon 34055, Republic of Korea}

\author[0000-0003-0562-5643]{Seung-Lee Kim}
\affiliation{Korea Astronomy and Space Science Institute, Daejon 34055, Republic of Korea}

\author[0000-0003-0043-3925]{Chung-Uk Lee}
\affiliation{Korea Astronomy and Space Science Institute, Daejon 34055, Republic of Korea}

\author{Dong-Joo Lee}
\affiliation{Korea Astronomy and Space Science Institute, Daejon 34055, Republic of Korea}

\author{Yongseok Lee}
\affiliation{Korea Astronomy and Space Science Institute, Daejon 34055, Republic of Korea}
\affiliation{School of Space Research, Kyung Hee University, Yongin, Kyeonggi 17104, Republic of Korea}

\author[0000-0002-6982-7722]{Byeong-Gon Park}
\affiliation{Korea Astronomy and Space Science Institute, Daejon 34055, Republic of Korea}
\affiliation{University of Science and Technology, Korea, (UST), 217 Gajeong-ro, Yuseong-gu, Daejeon 34113, Republic of Korea}

\author[0000-0003-1435-3053]{Richard W. Pogge}
\affiliation{Department of Astronomy, The Ohio State University, 140 W. 18th Avenue, Columbus, OH 43210, USA}

\collaboration{21}{(The KMTNet Collaboration)}

\author[0000-0001-7303-914X]{Uffe G. Jørgensen}
\affiliation{Centre for ExoLife Sciences, Niels Bohr Institute, University of Copenhagen, Øster Voldgade 5, 1350 Copenhagen, Denmark}

\author{Penélope Longa-Peña}
\affiliation{Centro de Astronomía, Universidad de Antofagasta, Avenida Angamos 601, Antofagasta 1270300, Chile}

\author[0000-0002-2859-1071]{Sedighe Sajadian}
\affiliation{Department of Physics, Isfahan University of Technology, Isfahan 84156-83111, Iran}

\author[0000-0003-1310-8283]{Jesper Skottfelt}
\affiliation{Centre for Electronic Imaging, Department of Physical Sciences, The Open University, Milton Keynes, MK7 6AA, UK}

\author[0000-0001-9328-2905]{Colin Snodgrass}
\affiliation{Institute for Astronomy, University of Edinburgh, Royal Observatory, Edinburgh, EH9 3HJ, UK}

\author[0000-0002-9024-4185]{Jeremy Tregloan-Reed}
\affiliation{Instituto de Investigación en Astronomía y Ciencias Planetarias, Universidad de Atacama, Copayapu 485, Copiapó, Atacama, Chile}

%
%
%
%
\author[0000-0002-8799-0080]{Nanna Bach-Møller}
\affiliation{Centre for Star and Planet Formation, Niels Bohr Institute, University of Copenhagen, Østervoldgade 5, DK-1350 Copenhagen, Denmark}

\author[0000-0002-5854-4217]{Martin Burgdorf}
\affiliation{Universität Hamburg, Faculty of Mathematics, Informatics and Natural Sciences, Department of Earth Sciences, Meteorological Institute, Bundesstraße 55, D-20146 Hamburg, Germany}

%
\author[0000-0001-9697-7331]{Giuseppe D'Ago}
\affiliation{Instituto de Astrofísica Pontificia Universidad Católica de Chile, Avenida Vicuna Mackenna 4860, Macul, Santiago, Chile}

%
%
\author[0000-0001-9279-2815]{Lauri Haikala}
\affiliation{Instituto de Astronomía y Ciencias Planetarias, Universidad de Atacama, Copayapu 485, Copiapó, Chile}

%
\author[0000-0002-1508-2243]{James Hitchcock}
\affiliation{University of St Andrews, Centre for Exoplanet Science, SUPA School of Physics \& Astronomy, North Haugh, St Andrews, KY16 9SS, UK}

\author[0000-0003-0961-5231]{Markus Hundertmark}
\affiliation{Zentrum für Astronomie der Universität Heidelberg, Astronomisches Rechen-Institut, Mönchhofstr. 12-14, D-69120 Heidelberg, Germany}

%
%
\author{Elahe Khalouei}
\affiliation{Astronomy Research Center, Research Institute of Basic Sciences, Seoul National University,1 Gwanak-ro, Gwanak-gu, Seoul 08826, Korea}

%
\author[0000-0002-6830-476X]{Nuno Peixinho}
\affiliation{Instituto de Astrof\'{\i}sica e Ci\^{e}ncias do Espa\c{c}o, Departamento de F\'{\i}sica, Universidade de Coimbra, PT3040-004 Coimbra, Portugal}

%
\author[0000-0002-7084-5725]{Sohrab Rahvar}
\affiliation{Department of Physics, Sharif University of Technology, PO Box 11155-9161, Tehran, Iran}

%
%
\author[0000-0002-3807-3198]{John Southworth}
\affiliation{Astrophysics Group, Keele University, Staffordshire, ST5 5BG, UK}

%
%
\author{Petros Spyratos}
\affiliation{Astrophysics Group, Keele University, Staffordshire, ST5 5BG, UK}

\collaboration{17}{(The MiNDSTEp Collaboration)}

\begin{abstract}
We present an analysis of microlensing event OGLE-2019-BLG-0825.
This event was identified as a planetary candidate by preliminary modeling.
We find that significant residuals from the best-fit static binary-lens model exist and a xallarap effect can fit the residuals very well and significantly improves $\chi^2$ values.
On the other hand, by including the xallarap effect in our models, we find that binary-lens parameters like mass-ratio, $q$, and separation, $s$, cannot be constrained well. 
However, we also find that the parameters for the source system like the orbital period and semi major axis are consistent between all the models we analyzed.
We therefore constrain the properties of the source system better than the properties of the lens system.
The source system comprises a G-type main-sequence star orbited by a brown dwarf with a period of $P\sim5$ days.
This analysis is the first to demonstrate that the xallarap effect does affect binary-lens parameters in planetary events.
It would not be common for the presence or absence of the xallarap effect to affect lens parameters in events with long orbital periods of the source system or events with transits to caustics, but in other cases, such as this event, the xallarap effect can affect binary-lens parameters.
\end{abstract}
\keywords{gravitational lensing: micro --- brown dwarfs --- xallarap}
\section{Introduction}
The gravitational microlensing method is a method for detecting exoplanets that utilizes the phenomenon that light is deflected by gravity \citep{Liebes1964,Paczynski1991} and is sensitive to planets beyond the snow line \citep{Gould+1992,Bennett+1996}.
Giant planets are thought to form near and beyond the snow line \citep{Ida+2004,Laughlin+2004,Kennedy+2006}.
In gravitational microlensing, when a lensing object crosses in front of a source star, the brightness of the source star changes with time owing to the gravitational effect of the lensing object. 
Furthermore, if this lensing object is accompanied by a planetary or binary-star companion, the gravity of this companion will cause a secondary magnification. 
The gravitational microlensing method does not use the light from the lensing object, but only the time-dependent variations arising form the gravitational effect of the lensing object or objects on the light from the source.
Therefore, the gravitational microlensing method has the advantage over other planet detection methods of being able to detect planets around faint stars at distances far from Earth \citep{Gaudi2012}. 
By comparing the occurrence rates of planets in the distant region detected by the gravitational microlensing method with the frequency of planets in the local region, we can investigate the influence of the Galactic environment on planet formation.

The detection of distant planets and brown dwarfs allows us to consider the influence of the Galactic environment on planet and brown dwarf formation.
It has been thought that different Galactic environments have different planetary occurrence rates \citep{Gonzalez+2001, Lineweaver+2004, Spinelli+2021}.
In fact, radial velocity surveys in the 25 pc region near the Sun reported that the occurrence rate of hot Jupiters is about $\sim2\%$ \citep{Hirsch+2021}, whereas Kepler transit surveys report that the occurrence rate of hot Jupiters around G- K- type stars near Cygnus is about $\sim0.5\%$ \citep{Howard+2012, Santerne+2012, Santerne+2016, Fressin+2013}. 
Although \citet{Koshimoto+2021a}  recently found that planetary frequencies do not depend significantly on the Galactocentric distance based on their 28 planet sample, their result is still too uncertain to discuss environmental effects precisely.

In the analysis of gravitational microlensing events, it is sometimes difficult to distinguish perturbations given by the lens secondary to the light curve from those given by higher-order effects \citep{Griest+1992,Rota+2021}. 
One of the higher-order effects, the parallax effect, is the effect of the acceleration of the Earth's orbital motion on the light curve. 
The xallarap effect is a higher-order effect on the light curve when the source is binary \citep{Griest+1992,Han+1997,Paczynski1997,Poindexter+2005}. 
Binary stars are common in the Universe, with binary systems of two or more stars accounting for about 30\% of all stellar systems \citep{Lada2006,Badenes+2018}. 
When a source is accompanied by a companion star, the companion is not necessarily magnified, but the light curve is affected by the orbital motion of the source primary \citep{Rota+2021}. 
Although most of the binary stars are too wide between their primary and companion stars to reliably detect a xallarap effect, a systematic survey of 22 long-term events in the bulge shows that 23\% of them are indeed affected by xallarap \citep{Poindexter+2005}. 
The effect of xallarap on lensing planet detection efficiency has not been fully investigated but is known to exist \citep{Zhu+2017}.

We present in this paper an analysis of OGLE-2019-BLG-0825 and report that the xallarap effect was detected and that the lensing system parameters changed before and after the xallarap effect was included.
Section~\ref{sec:Observation} describes the data for event OGLE-2019-BLG-0825.
Section~\ref{sec:Data_reduction} describes our data reduction.
Section~\ref{sec:light_curve_modeling} describes our modeling in detail. 
Section~\ref{sec:Source_Lens_Properties} derives the color and magnitude of the source and calculates the physical parameters of the source system from the color and magnitude of the source and the fitting parameters of the microlensing.
Section~\ref{sec:Lens_Properties} describes the estimation of the physical parameters of the lens system by Bayesian analysis. 
Finally, Section~\ref{sec:Discussion_and_Conclusion} discusses and summarizes the results of our analysis.
\section{Observation}
\label{sec:Observation}
\begin{figure*}
    \centering
    \includegraphics[scale=0.7]{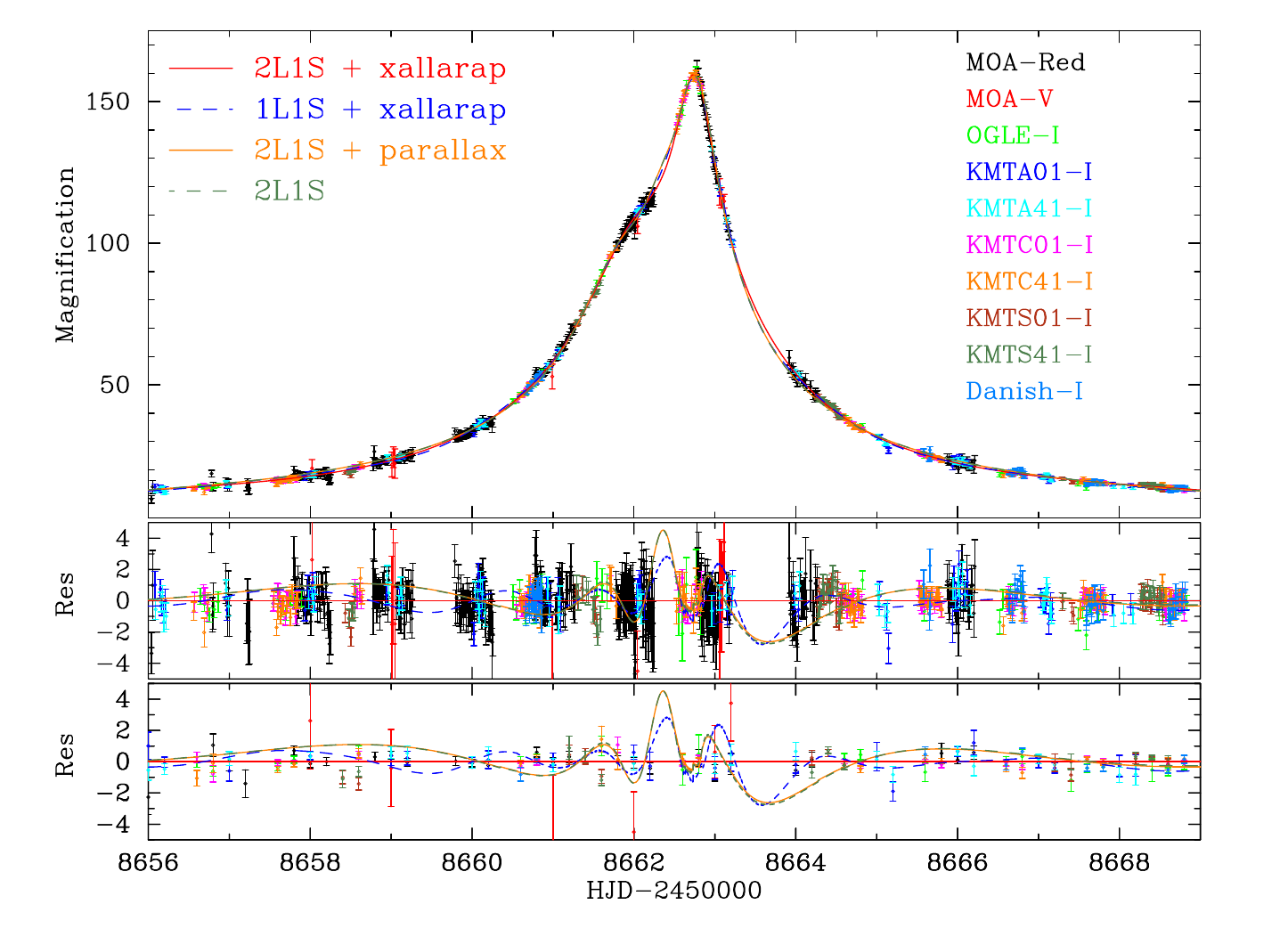}
    \caption{(Top panel) Light curve for OGLE-2019-BLG-0825. 
    Error bars are renormalized according to Equation (\ref{eq:error}).
    The red solid, blue dashed, orange solid, and green dashed lines are the best 2L1S + xallarap model, the best 1L1S + xallarap model, the best 2L1S + parallax model and the best standard 2L1S model described in Section~\ref{sec:light_curve_modeling}, respectively. 
    (Middle panel) Residuals from the best 2L1S + xallarap model.
    (Bottom panel) Residuals from the best 2L1S + xallarap model binned by 0.2 days.
    }
    \label{fig:lightcurve}
\end{figure*}
\begin{deluxetable*}{lccccccccccc}[t!]
\tablecaption{Data Sets for OGLE-2019-BLG-0825 \label{tab:dataset}}
\tablewidth{0pt}
\tablehead{
\multicolumn{1}{l}{Observatory Sites} & \multicolumn{1}{c}{Telescope} & \multicolumn{1}{c}{Collaboration} & \multicolumn{1}{c}{Label} & \multicolumn{1}{c}{Filter} & \multicolumn{1}{c}{$N_{\rm use}$} & \multicolumn{1}{c}{$k$\tablenotemark{1}} & \multicolumn{1}{c}{$e_{\rm min}$\tablenotemark{1}}
}
\startdata
Mount John & MOA-II 1.8m & MOA & MOA & MOA-Red & 3949 & 1.330 & 0.009 \\
& & & & $V$ & 86 & 0.835 & 0 \\
Las Campanas & Warsaw 1.3m & OGLE & OGLE & $I$ & 1535 & 1.453 & 0.007 \\
Siding Spring & KMTNet Australia 1.6m & KMTNet & KMTA01\tablenotemark{2} & $I$ & 704 & 1.649 & 0 \\
 & & & KMTA41\tablenotemark{3} & $I$ & 719 & 1.613 & 0 \\
Cerro Tololo Inter-American & KMTNet Chile 1.6m & & KMTC01\tablenotemark{2} & $I$ & 952	& 0.761	& 0.004 \\
& & & KMTC41\tablenotemark{3} & $I$ & 954 & 1.436 & 0 \\
South Africa Astronomical & KMTNet South Africa 1.6m & & KMTS01\tablenotemark{2} & $I$ & 881 & 1.490 & 0 \\
& & & KMTS41\tablenotemark{3} & $I$ & 887 & 1.416 & 0 \\
ESO’s La Silla & Danish 1.54m & MiNDSTEp & Danish & $I$ & 76 & 0.706 &	0 \\
\enddata
\tablenotetext{1}{Parameters for the error normalization.}
\tablenotetext{2}{Data observed in BLG01 in the overlapped area.}
\tablenotetext{3}{Data observed in BLG41 in the overlapped area.}
\end{deluxetable*}
Microlensing event OGLE-2019-BLG-0825 was first discovered on June 3, 2019 (${\rm HJD}^{\prime}\sim8638$)\footnote{${\rm HJD}^{\prime}\equiv{\rm HJD}-2450000$} at J2000 equatorial coordinates $({\rm R.A., decl.})=(17^h 52^m 21^s.62, -30^\circ 48^\prime 13^{\prime\prime}.2)$ corresponding to Galactic coordinates $(l,b)=(-0.849,-2.214)$, by the Optical Gravitational Lensing Experiment \citep[OGLE;][]{Udalski2003} collaboration. 
OGLE conducts a microlensing survey using the 1.3m Warsaw Telescope with a 1.4 deg$^2$ field-of-view (FOV) CCD camera at Las Campanas Observatory in Chile and distributes alerts of the discovery of microlensing events by their OGLE-IV Early Warning System \citep{Udalski+1994,Udalski2003,Udalski+2015}. 
The event is located in the OGLE-IV field BLG534, which is observed on Cousins $I$-band with an hourly cadence \citep{Mroz+2019}.

The Microlensing Observations in Astrophysics \citep[MOA;][]{Bond+2001,Sumi+2003} collaboration also independently discovered this event on June 23, 2019, and identified it as MOA-2019-BLG-273 using the MOA alert system \citep{Bond+2001}. 
The MOA collaboration conducts a microlensing exoplanet survey toward the Galactic bulge using the 1.8m MOA-II telescope with a 2.2 deg$^2$ wide FOV CCD camera, MOA-cam3 \citep{Sako+2008}, at the University of Canterbury's Mount John Observatory in New Zealand. 
The MOA survey uses a custom wide band filter referred to as $R_{\rm MOA}$ corresponding to the sum of the Cousins $R$ and $I$ bands.
In addition, a Johnson $V$-band filter is used primarily for measuring the color of the source. The event is located in the MOA field gb4, which is observed with high cadence once every 15 minutes.

The Korea Microlensing Telescope Network \citep[KMTNet;][]{Kim+2016} collaboration conducts a microlensing survey using three 1.6m telescopes each with a 4.0 deg$^2$ FOV CCD camera. 
The telescopes are located at the Cerro Tololo Inter-American Observatory (CTIO) in Chile, the South African Astronomical Observatory (SAAO) in South Africa, and Siding Spring Observatory (SSO) in Australia. 
This event is located in an overlapping region with two KMTNet observed fields (KMTNet BLG01 and BLG41), which are observed with high cadence once every 15 minutes and was discovered by the KMTNet EventFinder \citep{Kim+2018} as KMT-2019-BLG-1389 on June 28, 2019.

The Danish telescope of MiNDSTEp (Microlensing Network for the Detection of Small Terrestrial Exoplanets) made follow-up observations in $I$-band. 
MiNDSTEp uses the 1.54m Danish Telescope at the European Southern Observatory at La Silla Observatory in Chile \citep{Dominik+2010}.
Data from the Spitzer space telescope \citep{Yee+2015} were also available, but these show no detectable signal and so are not used.
A summary of all datasets used in the analysis of OGLE-2019-BLG-0825 is shown in Table~\ref{tab:dataset}.

The above data sets are used in our light curve analysis. 
To reduce long-term systematics on the baseline, we used approximately 2 years of data over $8100\leq {\rm HJD}^{\prime} \leq8800$.
Figure~\ref{fig:lightcurve} shows the light curve of OGLE-2019-BLG-0825 and the standard binary lens single source model (hereafter, standard 2L1S), the binary lens single source with parallax effect model (hereafter, 2L1S + parallax), the single lens single source with xallarap effct model (hereafter 1L1S + xallarap), and the best-fit model (binary lens single source with xallarap effect model, hearafter 2L1S + xallarap), described in Section~\ref{sec:light_curve_modeling}, respectively.
As will be discussed in detail in Section~\ref{sec:Source_Lens_Properties}, the xallarap model analysis assumes that the magnified flux of the second source is too weak to be detected, so it is denoted 1S.
\section{Data Reduction}
\label{sec:Data_reduction}
The OGLE data were reduced with the OGLE Difference Image Analysis (DIA) \citep{Wozniak2000} photometry pipeline \citep{Udalski2003,Udalski+2015} which uses the DIA technique \citep{Tomaney+1996,Alard+1998,Alard+2000}.  
The MOA data were reduced with MOA’s implementation of the DIA photometry pipeline \citep{Bond+2001}. 
The KMTNet data were reduced with their PySIS photometry pipeline \citep{Albrow+2009}. 
The MiNDSTEp data were reduced using DanDIA \citep{Bramich2008,Bramich+2013}. 

It is known that the nominal error bars calculated by the pipelines are incorrectly estimated in such crowded stellar fields. 
We follow a standard empirical error bar normalization process \citep{Yee+2012} intended to estimate proper uncertainties for the lensing parameters in the light-curve modeling. 
This process, described below, hardly affects the best-fit parameters \citep{Ranc+2019}.
We renormalize the photometric error bars using the formula
\begin{equation}\label{eq:error}
    \sigma^{\prime}_i = k\sqrt{\sigma^2_i + e^2_{\rm min}},
\end{equation}
in which $\sigma^{\prime}_i$ is the renormalized uncertainty in magnitude, while $\sigma_i$ is an uncertainty of the $i$-th original data point obtained from the photometric pipeline.
 The variables $k$ and $e_{\rm min}$ are renormalizing parameters. 
 For preliminary modeling, we search for the best-fit lensing parameters using $\sigma_i$.
 We then construct a cumulative $\chi^2$ distribution as a function of lensing magnification. 
 The $e_{\rm min}$ value is chosen so that the slope of the distribution is uniform \citep{Yee+2012}. 
 The $k$ value is chosen so that $\chi^2/$d.o.f.\footnote{Degrees of freedom.}$\simeq1$. 
 In Table~\ref{tab:dataset}, we list the calculated error bar renormalization parameters.
\section{Light Curve Modeling}

The model flux for a microlensing event is given by the following equation,
\begin{equation}\label{eq:magnification}
    f(t)=A(t,\bm{x})f_s + f_b,
\end{equation}

\noindent where $A(t,\bm{x})$ is the source flux magnification, $f_s$ is the flux of the source star, and $f_b$ is the blend flux.
In the 1L1S model, $\bm{x}$ is described by four parameters \citep{Paczynski1986}: the time of the source closest to the center of mass, $t_0$; the Einstein radius crossing time, $t_{\rm E}$; the impact parameter, $u_0$, and the source angular radius, $\rho$.
Both $u_0$ and $\rho$ are in units of the angular Einstein radius, $\theta_{\rm E}$.

For modeling the light curve, we used the Metropolis-Hastings Markov Chain Monte Carlo method. 
The finite source effect, an effect in which the source has a finite angular size, was calculated using the image-centered inverse-ray shooting method \citep{Bennett+1996,Bennett2010a} as implemented by \citet{Sumi+2010}. 
Note that $f_s$ and $f_b$ parameters are obtained from a linear-fit using the method of \citet{Rhie+1999}.
We adopt the following linear limb-darkening law for source brightness:
\label{sec:light_curve_modeling}
\begin{equation}
    S_\lambda(\vartheta) = S_\lambda(0)\left[1-u_\lambda(1-\cos(\vartheta))\right],
\end{equation}
where $\vartheta$ represents the angle between the line of sight and the normal to the surface of the source star. 
$S_\lambda(\vartheta)$ is a limb-darkening surface brightness of $\vartheta$ at wavelength $\lambda$. 
We estimated the effective temperature of the source star in Section~\ref{sec:Source_Lens_Properties} to be $T_{\rm eff}=5425\pm359$ K \citep{Gonzalez+2009}.
In this analysis, we assume the source star's metallicity $[\rm M/H]=0$, surface gravity $\log{g}=4.5$, and microturbulent velocity $v=1$ $\rm {km/s}$.
We use the limb-darkening cofficients $u_V=0.685$, $u_R=0.604$ and $u_I=0.518$, which are taken from the ATLAS model with $T_{\rm eff}=5500$ K \citep{Claret+2011}. 
Since $R_{\rm MOA}$ covers both $R$- and $I$-band wavelengths, we adopted the average value $u_{R_{\rm MOA}}=(u_R+u_I)/2=0.561$.
In addition, as will be discussed in more detail in Section~\ref{sec:Discussion_and_Conclusion}, we assume that the source of this event is a main-sequence star.
 
As the result of 1L1S model analysis, we found that $(t_0,t_{\rm E},u_0,\rho)=(8662.6, 47.6 ,1.2\times10^{-2},4.8\times10^{-3})$ is the best solution. 
This 1L1S best model is $\Delta\chi^2= 21400$ worse than the best standard 2L1S model.

\begin{figure*}
    \centering
    \includegraphics[scale=0.65]{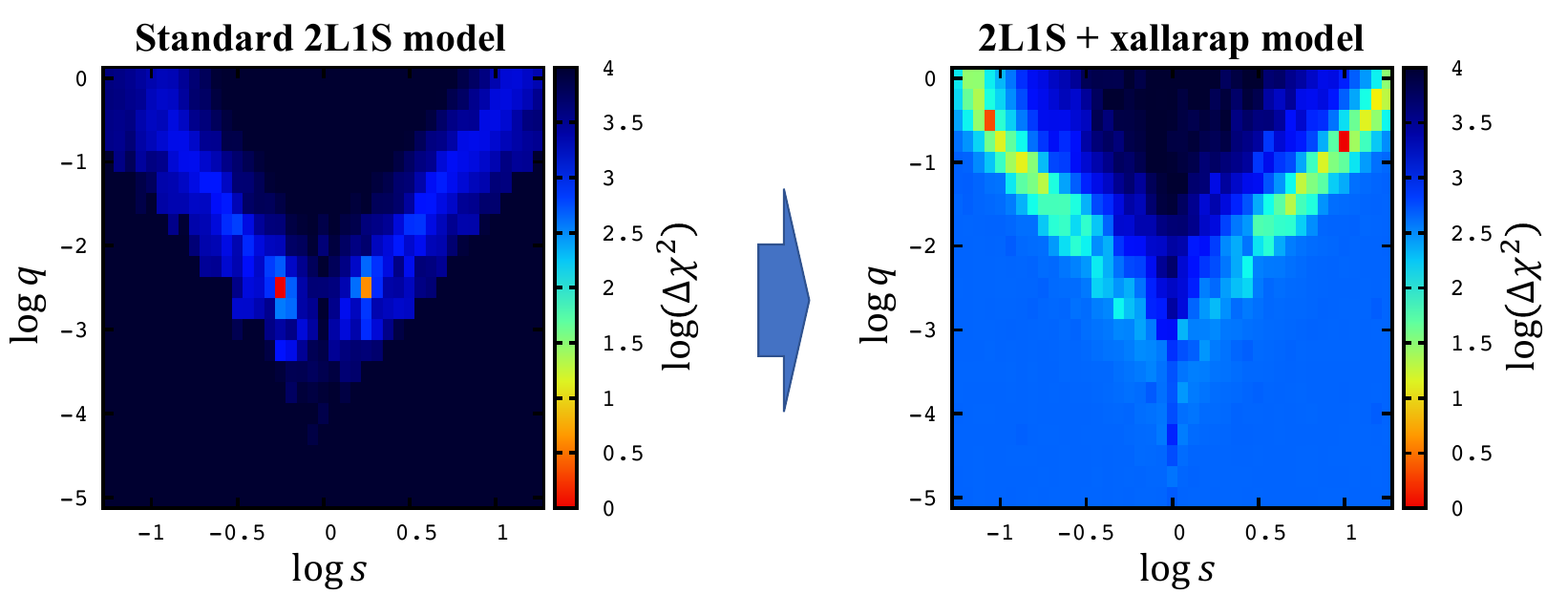}
    \caption{
    Map of $\Delta\chi^2$ in each $s–q$ grid from the $(q,s,\alpha)$ grid search for the standard 2L1S model (Left) and for the 2L1S + xallarap model (Right). 
    The best fit $\alpha$ is chosen for each grid location, respectively.
    In the map of the standard 2L1S model, we found the best solution at $q\sim10^{-3}$.
    However, for the 2L1S + xallarap map, best solutions at two other local minima appear at $q > 0.1$.
    }
    \label{fig:Grid_search}
\end{figure*}

\subsection{Standard Binary Lens}
\label{subsec:Standard_Binary_Lens}
In the standard 2L1S model, three additional parameters are required; the mass ratio of a lens companion relative to the host, $q$; the projected separation normalized by Einstein radius between binary components, $s$; the angle between the binary-lens axis and the source trajectory direction, $\alpha$.

Because the $\chi^2$ surface of the microlensing parameter has a very complicated shape, 34440 values of $(q,s,\alpha)$, which have a particularly large impact on the shape of the light curve were initially fixed in the fitting process.
Here we uniformly take 21 values between $-5\leq\log{q}\leq0$, 41 values between $-1.25\leq\log{s}\leq1.25$, 40 values in $0\leq\alpha\leq2\pi$, respectively.
For the top 1000 combinations which gave good fits, we performed the fitting again with $q,s$, and $\alpha$ free. 
This process minimizes the chance that we miss local solutions even in a large and complex microlensing parameter space. 
The left panel of Figure~\ref{fig:Grid_search} shows the results of the grid search analysis for the standard 2L1S model.

As a result of the analysis, the best fit standard 2L1S model is $(q,s)=(3.3\times10^{-3},0.57)$ (close1).
Hereafter, we call solutions with $s<1$ and $s>1$ as  ``close" and ``wide", respectively. We call the best standard 2L1S as close1.
We also found local minima at $(q,s)=(3.4\times10^{-3},1.75)$ (wide1) with $\Delta\chi^2\sim0.4$, $(q,s)=(2.1\times10^{-2},0.28)$ (close2) with $\Delta\chi^2\sim20.4$ and $(q,s)=(2.1\times10^{-2},3.78)$ (wide2) with $\Delta\chi^2\sim23.3$.
Detailed parameters of the standard binary models are shown in Table~\ref{tab:param_standard}. 
However, we observed systematic residuals around the peak of $8657<{\rm HJD}^\prime<8667$ in these models, as depicted by the green dashed line in Figure~\ref{fig:lightcurve}.
In Figure~\ref{fig:lightcurve}, we plot only close1, the best for the standard 2L1S, but the other three models also have similar residuals.
We therefore proceed to model the light curve with higher order effects.

\begin{deluxetable*}{c|ccccccccccc}[t!]
\tablecaption{Parameters of the standard 2L1S models \label{tab:param_standard}}
\tablewidth{0pt}
\tablehead{
\multicolumn{1}{c|}{Model} & \multicolumn{1}{c}{close1} & \multicolumn{1}{c}{close2} & \multicolumn{1}{c}{wide1} & \multicolumn{1}{c}{wide2} 
}
\startdata
$t_0$(HJD-2458660) & $2.474\pm0.001$ & $2.483\pm0.001$ & $2.473\pm0.001$ & 	$2.489\pm0.001$\\
$t_{\rm E}$ (days) & $74.7\pm2.0$ & $75.7\pm2.0$ & $72.8\pm1.8$ & $77.3\pm2.0$ \\
$u_0$ ($10^{-3}$) & $7.30\pm0.21$ &  $7.11\pm0.19$ & $7.53\pm0.19$ & $6.91\pm0.19$\\
$q$ ($10^{-3}$) & $3.30\pm0.11$ & $20.71\pm9.84$ & $3.39\pm0.10$ & $21.33\pm1.15$  \\
$s$ & $0.569\pm0.004$ & $0.207\pm0.038$ & $1.747\pm0.011$ & 
 $3.776\pm0.063$ \\
$\alpha$ (radian) & $5.034\pm0.002$ & $2.766\pm0.002$ & $5.036\pm0.003$ & $2.767\pm0.002$\\
$\rho$ ($10^{-3}$) & $2.95\pm0.09$ & $0.48\pm0.28$ & $3.02\pm0.09$ & $0.47\pm0.14$\\ \hline \hline
$\chi^2$ & 11744.7 & 11765.1 & 11745.1 & 11768.0\\
$\Delta\chi^2$ & - & 20.4 &	0.4 & 23.3\\
\enddata
\end{deluxetable*}
\begin{figure*}
    \centering
    \includegraphics[scale=0.7]{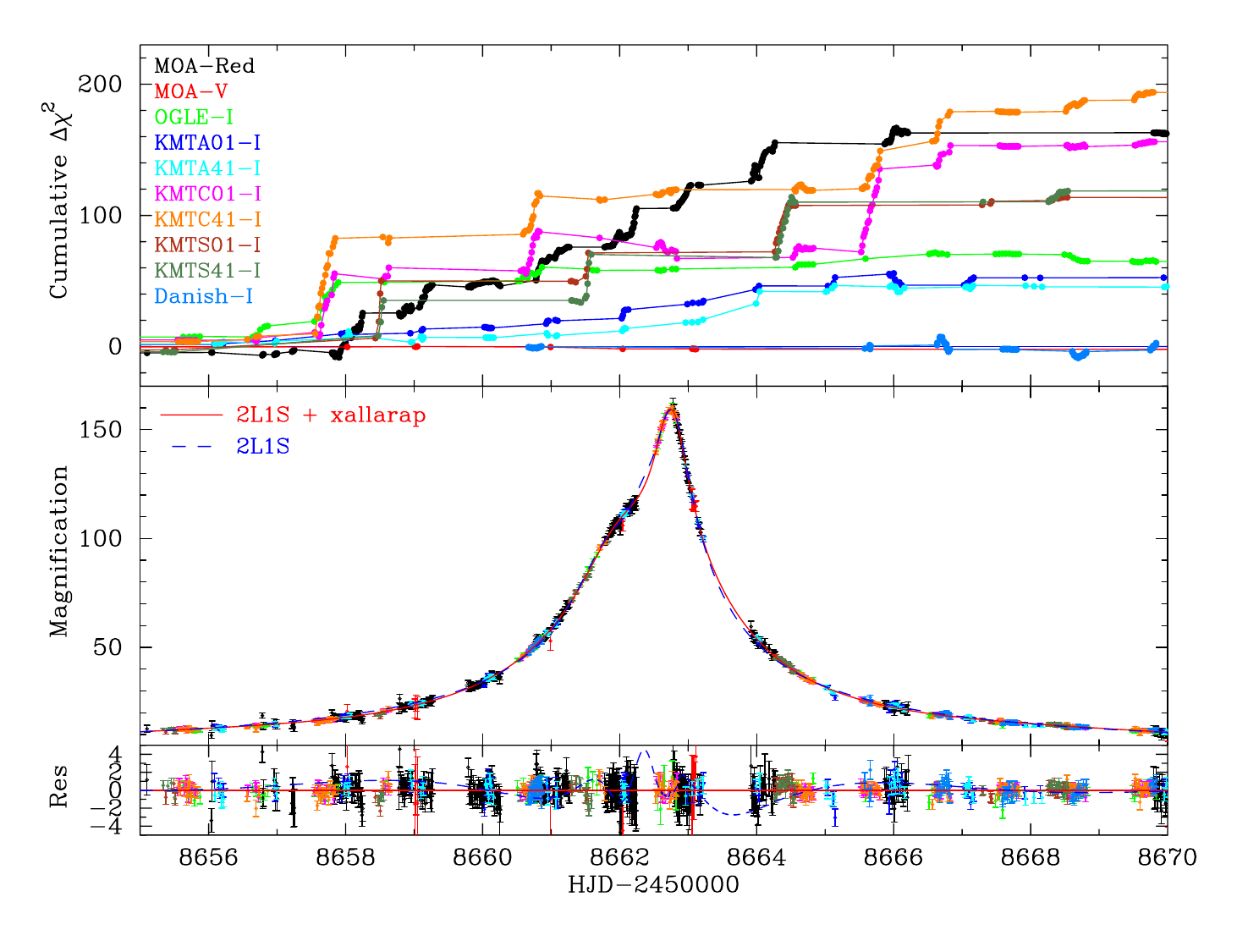}
    \caption{
    (Top panel) Cumulative $\Delta \chi^2$ for the xallarap model compared to the standard binary lens model. 
    Each color corresponds to each instrument listed the left side of the axes. 
    (Middle panel) The light curve of the best 2L1S + xallarap model (solid red line), and the light curve of the standard 2L1S best model (blue dashed line). 
    (Bottom panel) Residuals of the light curves from the 2L1S + xallarap model.
    }
    \label{fig:lc3}
\end{figure*}
\begin{deluxetable*}{c|ccccccccccc}[t!]
\tablecaption{\textbf{Parameters of the 2L1S + xallarap models, 1L1S + xallarap model, and 1L1S + xallarap + parallax model}
\label{tab:param_xallarap}}
\tablewidth{0pt}
\tablehead{Model & XLclose1 & XLclose2 & XLwide1 & XLwide2 & 1LXL & 1LXLPL\\
\multicolumn{1}{c|}{range of $q$} & \multicolumn{1}{c}{$q\leq0.1$} & \multicolumn{1}{c}{$0.1<q\leq1$} & \multicolumn{1}{c}{$q\leq0.1$} & \multicolumn{1}{c}{$0.1<q\leq1$} & - & -
}
\startdata
$t_0$(HJD-2458660) & $2.576\pm0.005$ & $2.573\pm0.007$ & $2.572\pm0.004$ & $ 2.575\pm0.006$ & $2.744\pm0.001 $ & $2.744\pm0.001$\\
$t_{\rm E}$ (days) & $97.7\pm2.6$ & $93.8\pm3.1$ & $100.7\pm3.4$ & $133.3\pm11.5$ & $ 67.2\pm2.0$ & $73.5\pm1.5$\\
$u_0$ ($10^{-3}$) & $-7.16\pm0.22$ & $-6.98\pm0.21$ & $7.06\pm0.21$ & $4.91\pm0.35$ & $ 6.93\pm0.20 $ & $6.31\pm0.14$\\
$q$ & $0.09\pm0.01$ & $0.44\pm0.11$ & $0.10\pm0.01$ & $0.94\pm0.37$ & - & -\\
$s$ & $0.141\pm0.004$ & $0.085\pm0.004$ & $7.403\pm0.438$ & $18.040\pm1.263$ & - & -\\
$\alpha$ (radian) & $0.429\pm0.008$ & $1.937\pm0.010$ & $5.846\pm0.008$ & $4.350\pm0.009$ & - & -\\
$\rho$ ($10^{-3}$) & $2.56\pm0.13$ & $2.15\pm0.12$ & $2.41\pm0.11$ & $1.44\pm0.16$ & $7.04 \pm 0.20$ & $6.41\pm0.14$\\
${\rm RA}_\xi$ (degree) & $81.6\pm11.7$ & $153.2\pm10.5$ & $75.9\pm14.5$  & $155.4\pm8.5$ & $ 31.3\pm0.5$ & $32.1\pm0.5$\\
${\rm decl}_\xi$ (degree) & $54.5\pm10.5$ & $36.9\pm16.4$ & $-79.4\pm12.8$ & $-40.7\pm14.1$ & $9.9\pm0.2$ & $9.9\pm0.3$\\
$P_\xi$ (days) & $5.42\pm0.04$ & $5.53\pm0.05$ & $5.43\pm0.04$ & $5.54\pm0.05$ & $2.91\pm0.02$ & $2.9\pm0.02$\\
$\xi_{{\rm E,N}}$ ($10^{-3}$) & $1.82\pm0.15$ & $-0.36\pm0.36$ & $-1.65\pm0.11$ & $0.34\pm0.20$ & $-3.59\pm0.11$ & $-3.23\pm0.08$\\
$\xi_{{\rm E,E}}$ ($10^{-3}$) & $0.69\pm0.34$ & $1.53\pm0.12$ & $0.42\pm0.41$ & $1.07\pm0.10$ & $2.86\pm0.09$ & $2.58\pm0.07$\\
$\pi_{{\rm E,N}}$ & - & - & - & - & - & $0.09\pm0.05$\\
$\pi_{{\rm E,E}}$ & - & - & - & - & - & $0.26\pm0.14$\\
\hline \hline
$\chi^2$ & 10856.4 & 10840.9 & 10861.2 & 10842.7 & 11311.5 & 11285.7\\
$\Delta\chi^2$ & 15.5 &	- & 20.3 & 1.8 & 470.6 & 444.8\\
\enddata
\end{deluxetable*}
\subsection{Parallax}
\label{subsec:Parallax}
It is known that the acceleration of Earth orbital motion affects the light curve of microlensing events \citep{Gould1992,Gould2004,Smith+2003,Dong+2009}.  
This parallax effect can be described by the microlensing parallax vector $\bm{\pi_{\rm E}} = (\pi_{{\rm E,N}}, \pi_{{\rm E,E}})$ where $\pi_{{\rm E,N}}$ and $\pi_{{\rm E,E}}$ represents respectively the north and east components of $\bm{\pi_{\rm E}}$ projected onto the sky plane in equatorial coordinates. 
The direction of $\bm{\pi_{\rm E}}$ is defined to coincide with the direction of the geocentric lens-source relative proper motion projected onto the sky plane at the reference time $t_{\rm fix}$, and the amplitude of $\bm{\pi_{\rm E}}$ is $\pi_{\rm E}={\rm au}/\tilde r_{\rm E}$ ($\tilde r_{\rm E}$ is the Einstein radius projected inversely to the observation plane) \citep{Gould2000}.

As a result of modeling by adding two parameters of $\pi_{{\rm E,N}}$ and $\pi_{{\rm E,E}}$, we found two degenerate models with $(q,s)=(3.5\times10^{-3},0.57)$ and $(q,s)=(3.4\times10^{-3},1.74)$, that are better than the standard 2L1S model by $\Delta\chi^2=68.3$.
However, the cumulative $\Delta\chi^{2}$ improvement for parallax model relative to standard 2L1S model is not consistent between the data sets.
Furthermore, we still found systematic residuals around the peak of $8657<{\rm HJD}^\prime<8667$ in these models, as seen in the standard 2L1S model shown by the orange solid line in Figure~\ref{fig:lightcurve}.

\subsection{Xallarap}
\label{subsec:Xallarap}
We next consider the possibility that the short term residuals in $8657<{\rm HJD}^\prime<8667$ are caused by a short period binary source system, i.e., they arise owing to the xallarap effect.

The xallarap effect can be described by the following seven parameters; the direction toward the solar system relative to the orbital plane of the source system, ${\rm RA}_\xi$ and ${\rm decl}_\xi$; the source orbital period, $P_\xi$; the source orbital eccentricity $e_\xi$; the perihelion $T_{\rm peri}$; the xallarap vector, $\bm{\xi_{\rm E}} = (\xi_{{\rm E,N}}, \xi_{{\rm E,E}})$. 
Note that this effect does not include the magnifying effect of the source companion star; only the source host contributes to the magnification.
We denote this model of the microlensing event as the 2L1S + xallarap model rather than as the 2L2S model to distinguish it from a model including secondary source magnification.
As discussed in detail in Section~\ref{sec:Source_Lens_Properties}, the flux ratio of the source companion to the host star in the $I$-band in the best 2L1S + xallarap model is $\sim10^{-7}$.
Therefore, we assume that the brightening of the source companion star is negligible.

We first fit using 78,960 values of xallarap parameters $({\rm RA}_\xi, {\rm decl}_\xi, P_\xi)$ with the four best standard 2L1S models (close1, wide1, close2, and wide2) as initial values.
We used 20 evenly spaced values for $0\leq{\rm RA}_\xi<360$, 21 values for $-90\leq{\rm decl}_\xi<90$, 19 and 99 values for $1<P_\xi$ [days] $<19$ and $20<P_\xi$ [days] $<1000$, respectively.
After that, we fit again with $({\rm RA}_\xi, {\rm decl}_\xi, P_\xi)$ as free parameters.
As a result, we found the best solutions with $P_\xi\sim5$ days independently from the initial values of close1, wide1, close2, and wide2.
We also found that the final $q$ and $s$ values are quite different from their initial values, and did not converge. 
Therefore, we next set $P_\xi\sim5$ days as the initial value, ${\rm RA}_\xi$ and ${\rm decl}_\xi$ to random values, and performed model fitting with 34,440 values of $(q,s,\alpha)$ using the same procedure as the standard 2L1S modeling described in Section~\ref{subsec:Standard_Binary_Lens}.
Short-period binary stars orbiting in $P_\xi\sim5$ days are affected by orbital circularization due to tidal friction \citep{Fabrycky+2007}.
The tidal circularization time is discussed in Section~\ref{sec:Discussion_and_Conclusion}, but it is reasonable to assume that at the age of the stars in the Galactic bulge \citep{Sit+2020}, the orbit is fully circularized.
Therefore, we fixed the eccentricity at $e_\xi=0$.
When $e_{\xi}=0$, $T_{\rm peri}$ can be eliminated as a fitting parameter.
The results are shown in the right panel of Figure~\ref{fig:Grid_search}. 

\begin{figure}
    \centering
    \includegraphics[scale=0.7]{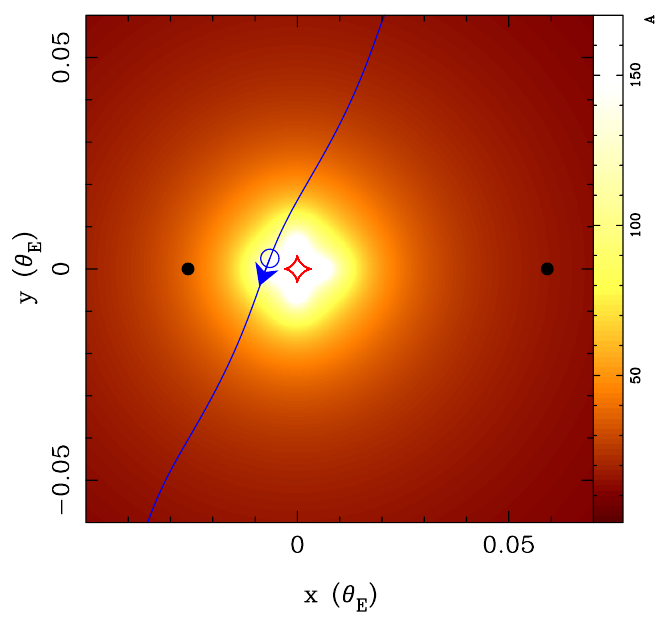}
    \caption{
    The geometry of the primary lens, source trajectory, caustics on the magnification map for the best 2L1S + xallarap model.
    The black filled circle on the left indicates the primary lens.
    The black filled circle on the right indicates the lens companion.
    The blue line with arrow represents the source trajectory.
    The blue circles represent the source size and position at $t_0$. 
    The red closed curve represents the caustic.
    The colored contours represent the magnification map. 
    }
    \label{fig:caustic}
\end{figure}

The figure shows that there are degenerate solutions for various combinations of $(q,s)$ values in the range of $\Delta\chi^2\lesssim20$.
Table~\ref{tab:param_xallarap} shows the best fit model parameters for the wide and close solutions.
The reason for the slight difference in $\Delta\chi^2$ between Figure~\ref{fig:Grid_search} and Table~\ref{tab:param_xallarap} is that the models in Table~\ref{tab:param_xallarap} were fitted with $q$, $s$, and $\alpha$ set free.
We label the best models of the mass ratio range in the 2L1S + xallarap close model, respectively: the best with $q\leq0.1$ is XLclose1, the best with $0.1<q\leq1$ is XLclose2.
Similarly, in the wide model of 2L1S + xallarap, we label the best with $q\leq0.1$ as XLwide1, the best with $0.1<q\leq1$ as XLwide2.
Figure~\ref{fig:lightcurve} shows the best 2L1S + xallarap model (i.e. XLclose2).
The xallarap models fit the light curves better than the standard 2L1S models.
Figure~\ref{fig:lc3} shows the cumulative $\Delta \chi^2$ of the best 2L1S + xallarap model relative to the best standard 2L1S model.
One can see that the 2L1S + xallarap model improves $\chi^2$ around the peak of $8657<{\rm HJD}^\prime<8667$.
The 2L1S + xallarap model improved $\chi^2$ by $903.7$ from the standard 2L1S model and by $835.5$ from the 2L1S + parallax model. 
Figure~\ref{fig:caustic} shows the geometry of the primary lens, source trajectory, caustics on the magnification map for the best 2L1S + xallarap model.
The short orbital period of the source star with $P_\xi\sim5$ days make the source's trajectory a wavy line.

We applied the same procedure for 1L1S and found the best 1L1S + xallarap model has $\Delta\chi^2=470.6$ worse than the best 2L1S + xallarap model.
We label the best 1L1S + xallarap model as 1LXL.
Even the 1L1S + xallarap + parallax model was $\Delta\chi^2=444.8$ worse than the best 2L1S + xallarap model.
We label the best 1L1S + xallarap + parallax model as 1LXLPL.
The parameters of each of the best models are listed in Table~\ref{tab:models}.
However, asymmetric maps similar to Figure~\ref{fig:caustic} can be created by binary lenses of various parameters, which led to the emergence of various degenerate 2L1S + xallarap models.
We considered other higher order effects and combinations of them such as 2L1S + xallarap + parallax, 2L1S + xallarap + parallax + lens orbital motion, and 1L2S, but could not detect them significantly.
For comparison with the 2L1S + xallarap model, we also fitted the 2L1S model with a variable source.
In this case, the amplitude of the variation, $\gamma$; the period of the variation, $T_{\rm v}$; and the initial phase, $\beta$ are additional parameters.
We fixed the other parameters at those of the best standard 2L1S model (i.e., close1).
However, the $\chi^2$ improvement from the best standard 2L1S model was only $139.1$, $\Delta \chi^2 = 764.6$ worse than the best 2L1S + xallarap model.
To confirm, we performed 2L1S + xallarap fitting analysis with $\xi_{\rm E,N}$, $\xi_{\rm E,E}$, ${\rm RA}_\xi$, ${\rm decl}_\xi$, and $P_\xi$ set free and the other parameters fixed to the best standard 2L1S model.
As a result, the $\chi^2$ was improved by $594.5$ over the best standard 2L1S model.
This is only $\Delta \chi^2 = 309.3$ worse than the best 2L1S + xallarap model.
That is, for two models (2L1S + xallarap and 2L1S + variable source) with the same fixed lens parameters, the 2L1S + xallarap model has $455.3$ better $\chi^2$ than the 2L1S + variable source model.
Finally, we conclude that the best model in this analysis is XLclose2.
In addition, the xallarap signal is consistent, and considering additional higher order effects on 2L1S + xallarap has little influence on our conclusions.
%
%

\begin{deluxetable}{ccccccccccc}[t!]
\tablecaption{Comparisons between each microlensing model
\label{tab:models}}
\tablewidth{0pt}
\tablehead{
\multicolumn{1}{c}{Model} & \multicolumn{1}{c}{$N_{\rm param}$} & \multicolumn{1}{c}{$\chi^2$} & \multicolumn{1}{c}{$\Delta\chi^2$}
}
\startdata
1L1S & 4 & 33144.7 & 22303.8\\
1L1S + xallarap & 9\tablenotemark{\rm *}& 11311.5 & 470.6\\
1L1S + xallarap + parallax & 11\tablenotemark{\rm *}& 11285.7 & 444.8\\
standard 2L1S & 7 & 11744.7 & 903.7\\
2L1S + parallax & 9 & 11676.5 & 835.5\\
2L1S + xallarap & 12\tablenotemark{\rm *}& 10840.9 & -\\
\enddata
\tablenotetext{*}{The source orbital eccentricity is fixed at $e_{\xi}=0$. 
When $e_{\xi}=0$, $T_{\rm peri}$ can be eliminated because it is a parameter that cannot take a specific value.}
\end{deluxetable}

\section{Source System Properties}
\label{sec:Source_Lens_Properties}
\begin{figure}
    \centering
    \includegraphics[scale=0.45]{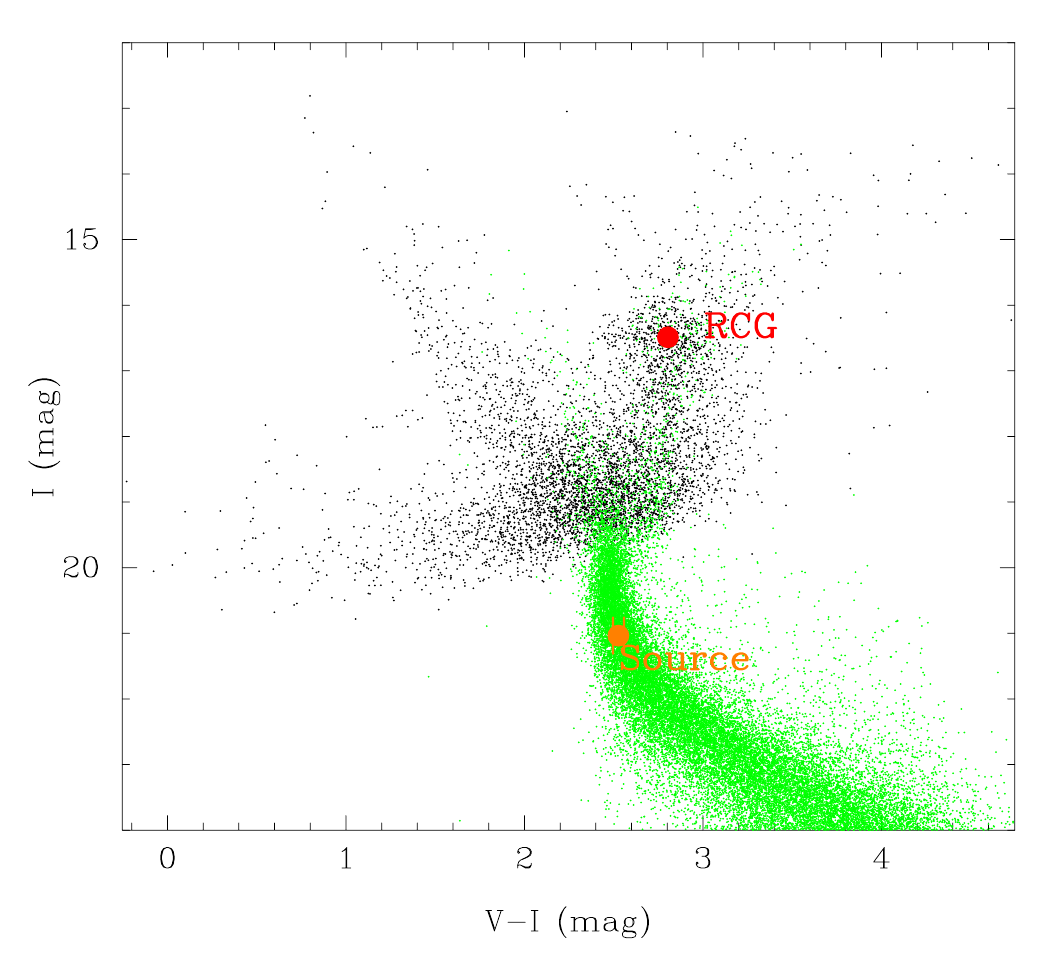}
    \caption{
    Color Magnitude Diagram (CMD, black dots) of the OGLE-$\mathrm{I}\hspace{-1.2pt}\mathrm{I}\hspace{-1.2pt}\mathrm{I}$ stars within $2’$ around OGLE-2019-BLG-0825. The green dots are stars in Baade's window based on Hubble Space Telescope observations \citep{Holtzman+1998}, color- and magnitude-matched at the RCG position. The orange circles represent the positions of the source, and the red dots represent the positions of the RCG centroid within $2’$ around OGLE-2019-BLG-0825.
    }
    \label{fig:cmd}
\end{figure}
We estimated the angular source radius, $\theta_*$, from the color and magnitude of the source.
The best fit instrumental source magnitudes of $R_{\rm MOA}$ and $V_{\rm MOA}$ are calibrated to the Cousins $I$-band and Johnson $V$-band magnitude scales by cross-referencing to the stars in the OGLE-$\mathrm{I}\hspace{-1.2pt}\mathrm{I}\hspace{-1.2pt}\mathrm{I}$ photometry map \citep{Szymanski+2011} within $0\farcs7$ of the event.

For reliability, we restricted stars to $16 \leq V_{\rm OGLE-\mathrm{I}\hspace{-1.2pt}\mathrm{I}\hspace{-1.2pt}\mathrm{I}}$ [mag] $\leq 19$, and performed $5\sigma$ clipping in the linear regressions of $V_{\rm MOA}$ vs. $V_{\rm OGLE-\mathrm{I}\hspace{-1.2pt}\mathrm{I}\hspace{-1.2pt}\mathrm{I}}$ and $(I_{\rm OGLE-\mathrm{I}\hspace{-1.2pt}\mathrm{I}\hspace{-1.2pt}\mathrm{I}}-R_{\rm MOA})$ vs. $(V-R)_{\rm MOA}$ and $(V-I)_{\rm OGLE-\mathrm{I}\hspace{-1.2pt}\mathrm{I}\hspace{-1.2pt}\mathrm{I}}$ vs. $(V-R)_{\rm MOA}$, respectively.
From the final 73 remaining objects, the following conversion equations from $R_{\rm MOA}$ and $(V-R)_{\rm MOA}$ to $I_{\rm OGLE-\mathrm{I}\hspace{-1.2pt}\mathrm{I}\hspace{-1.2pt}\mathrm{I}}$ and $(V-I)_{\rm OGLE-\mathrm{I}\hspace{-1.2pt}\mathrm{I}\hspace{-1.2pt}\mathrm{I}}$ were obtained by linear regression,

\begin{equation}\label{eq:convert1}
    I_{\rm OGLE-\mathrm{I}\hspace{-1.2pt}\mathrm{I}\hspace{-1.2pt}\mathrm{I}} = R_{\rm MOA}-(0.24\pm0.01) \times (V-R)_{\rm MOA} + (27.22\pm0.01),
\end{equation}

\begin{equation}\label{eq:convert2}
    (V-I)_{\rm OGLE-\mathrm{I}\hspace{-1.2pt}\mathrm{I}\hspace{-1.2pt}\mathrm{I}} = (1.20\pm0.01) \times (V-R)_{\rm MOA} + (0.94\pm0.02).
\end{equation}

As a result, the color and magnitude with the extinction of the source star for the best fit 2L1S + xallarap model  was $(V-I,I)_{\rm S}=(2.527 \pm 0.031,$ $21.035 \pm 0.015)$.
The intrinsic color and magnitude of RCG stars are $(V-I,I)_{{\rm RCG},0}=(1.060 \pm 0.060,$ $14.443 \pm 0.040)$ \citep{Bensby+2013,Nataf+2013}.
From the color-magnitude diagram of the stars within $2'$ of the source star (Figure~\ref{fig:cmd}), the RCG centroid is estimated as $(V-I,I)_{\rm RCG}=(2.804 \pm 0.009,$ $16.488 \pm 0.022)$.
Then we calculated $(E(V-I),A(I))=(1.744 \pm 0.061,$ $2.045 \pm 0.046)$.
Finally, we have the intrinsic color and magnitude of the source star $(V-I,I)_{\rm S,0}=(0.783 \pm 0.068,$ $18.990 \pm 0.048)$ for the best 2L1S + xallarap model.
Also, Figure \ref{fig:cmd} shows that the source is a main-sequence star and unlikely to be a variable star.
Table~\ref{tab:Source_xallarap} shows that the values for $(V-I,I)_{\rm S,0}$ for the other models are almost the same.
%

We estimated the angular source radius of $\theta_* = 0.538 \pm 0.039$ $\mu$as from the relation,
\begin{equation}\label{eq:theta_star}
    \log(2\theta_*)=0.50 + 0.42(V-I)_0 - 0.2 I_0,
\end{equation}
where the accuracy of the relational equation is better than 2\% \citep{Fukui+2015}.
This relation is based on \citet{Boyajian+2014}, but derived by  limiting to FGK stars with $3900$ $<T_{\rm {eff}}$ [K] $<7000$ (Boyajian, 2014, Private Communication).
Then, we calculated the lens's Einstein radius of $\theta_{\rm E}= \rho \theta_* = 0.25 \pm 0.02$ mas and the lens-source relative proper motion of $\mu_{\rm rel}=\theta_{\rm E}/t_{\rm E} = 0.97 \pm 0.10$ mas $\rm {yr^{-1}}$.

The amplitude of the xallarap vector, $\xi_{\rm E}$ is described as follows,
\begin{equation}\label{eq:xi_E}
    \xi_{\rm E}\equiv \Bigl (\frac{\theta_{\rm E} D_{\rm S}}{1\:\rm{au}} \Bigr)^{-1} \Bigl (\frac{P_\xi}{1\: \rm {yr}} \Bigr )^{2/3} \Bigl (\frac{M_{\rm S,C}}{M_\odot} \Bigr ) \Bigl (\frac{M_{\rm S,H}+M_{\rm S,C}}{M_\odot} \Bigr )^{-2/3},
\end{equation}

\noindent where $M_{\rm S,H}$ and $M_{\rm S,C}$ are the masses of host and companion of the source system, respectively.
$M_{\rm S,H}$ is estimated by using isochrones \citep[PARSEC;][]{Bressan+12} and the absolute magnitude of the host source star $M(I_{\rm S})=I_{\rm {S,0}} + 5 \log_{10}{D_{\rm S} {\rm {[pc]}}}+5=4.48\pm0.38$ mag assuming $D_{\rm S}=8.0\pm1.4$ kpc.
Then, $M_{\rm S,C}$ can be solved from Equation (\ref{eq:xi_E}).
Also, using Kepler's third law, 

\begin{equation}\label{eq:Kepler}
    \Bigl (\frac{a_{\rm S}}{1\:\rm {au}} \Bigr ) ^3 \Bigl (\frac{P_\xi}{1\:\rm {yr}} \Bigr ) ^{-2} = \Bigl [\frac{M_{\rm S,H} + M_{\rm S,C}}{M_{\odot}} \Bigr],
\end{equation}

\noindent we can solve $a_{\rm S}$ which is the semi-major axis of the source system. 
The apparent $H$- and $K$-bands magnitudes of the source with extinction $H_{\rm S}$ and $K_{\rm S}$ are also estimated using PARSEC isochrones and the wavelength dependence of the extinction law in the direction of Galactic center, $A_V:A_H:A_{K_s}=1:0.108:0.062$ \citep{Nishiyama+2008}.
In addition, we calculated $L_{\rm S,C}/L_{\rm S,H}$, the luminosity ratio in the $I$-band of the source companion $L_{\rm S,C}$ to the source host $L_{\rm S,H}$.
For this we used the mass-luminosity empirical relation of \citet{Bennett+2015}, which combines \citet{Henry+1993} and \citet{Delfosse+2000}, and the isochrone model of \citet{Baraffe+2003}.
We used the \citet{Henry+1993} relation for $M>0.66$ $M_{\odot}$, the \citet{Delfosse+2000} relation for $0.12$ $M_{\odot}<M<0.54$ $M_{\odot}$.
For low-mass stars ($M<0.10$ $M_{\odot}$) we used the isochrone model of \citet{Baraffe+2003} for sub-stellar objects at an age of 10 Gyr.
At the boundary of these mass ranges, we interpolated linearly between the two relations.
Table~\ref{tab:Source_xallarap} shows our calculated properties of the source system for the 2L1S + xallarap models in Table~\ref{tab:param_xallarap}.
The source host in the best 2L1S + xallarap model is a G-type main-sequence star and the source companion is a brown dwarf with a semi major axis of $a_{\rm S}=0.0594\pm0.0005$ au.
The luminosity ratio at the $I$-band of the source companion $L_{\rm S,C}$ is small, $L_{\rm S,C}/L_{\rm S,H}=(1.0\pm0.3)\times10^{-7}$, and does not conflict with our assumption the magnified flux of the second source is too weak to be detected.

\begin{deluxetable*}{c|ccccccccccc}[t!]
\tablecaption{Source system properties of the 2L1S + xallarap models 
\label{tab:Source_xallarap}}
\tablewidth{0pt}
\tablehead{Model & XLclose1 & XLclose2 & XLwide1 & XLwide2 \\
\multicolumn{1}{c|}{range of $q$} & \multicolumn{1}{c}{$q\leq0.1$} & \multicolumn{1}{c}{$0.1<q\leq1$} & \multicolumn{1}{c}{$q\leq0.1$} & \multicolumn{1}{c}{$0.1<q\leq1$}
}
\startdata
$V_{\rm S}$ (mag) & $23.58\pm0.03$ & $23.56\pm0.03$ & $23.55\pm0.03$ & $23.56\pm0.03$\\
$I_{\rm S}$ (mag) & $21.06\pm0.01$ & $21.04\pm0.01$ & $21.02\pm0.01$ & $21.06\pm0.01$\\
$H_{\rm S}$ (mag) & $18.54\pm0.30$ & $18.52\pm0.484$ & $18.51\pm0.48$ & $18.54\pm0.48$\\
$K_{\rm S}$ (mag) & $18.31\pm0.48$ & $18.29\pm0.48$ & $18.28\pm0.48$ & $18.31\pm0.48$\\
$(V-I)_{\rm S}$ (mag) & $2.52\pm0.03$ & $2.53\pm0.03$ & $2.52\pm0.03$ & $2.527\pm0.03$\\
$I_{\rm S,0}$ (mag) & $19.01\pm0.05$ & $18.99\pm0.05$ & $19.00\pm0.05$ & $19.01\pm0.05$\\
$(V-I)_{\rm S,0}$ (mag) & $0.78\pm0.07$ & $0.78\pm0.07$ & $0.78\pm0.07$ & $0.78\pm0.07$\\
$M_I$ (mag) & $4.50 \pm 0.38$ & $4.48 \pm 0.38$ & $4.46 \pm 0.38$ & $4.50 \pm 0.38$\\
$\theta_{\rm E}$ (mas) & $0.53\pm0.04$ & $0.25\pm0.02$ & $0.22\pm0.02$ & $0.37\pm0.05$\\
$\mu_{\rm rel}$ (mas ${\rm yr}^{-1}$) & $0.78\pm0.07$ & $0.97\pm0.10$ & $0.81\pm0.07$ & $1.01\pm0.16$\\
$M_{\rm S,H}$ ($M_\odot$) & $0.864\pm0.045$ & $0.867\pm0.045$ & $0.868\pm0.045$  & $0.864\pm0.045$\\
$M_{\rm S,C}$ ($M_\odot$) & $0.050\pm0.005$ & $0.048\pm0.004$ & $0.047\pm0.004$ & $0.051\pm0.006$\\
$a_{\rm S}$ ($10^{-2}$ au) & $5.86\pm0.04$ & $5.94\pm0.05$ &  $5.87\pm0.03$ & $5.95\pm0.05$\\
$L_{\rm S,C}/L_{\rm S,H}$ ($10^{-7}$) & $1.15\pm0.34$ &  $1.02\pm0.26$ & $0.95\pm0.23$ & $1.21\pm0.47$\\
\hline \hline
$\chi^2$ & 10856.4 & 10840.9 & 10861.2 & 10842.7\\
$\Delta\chi^2$ & 15.5 & - & 20.3 & 1.8\\
\enddata
\end{deluxetable*}

\section{Lens System Properties by Bayesian Analysis}
\label{sec:Lens_Properties}

\begin{figure*}
    \centering
    \includegraphics[scale=0.69]{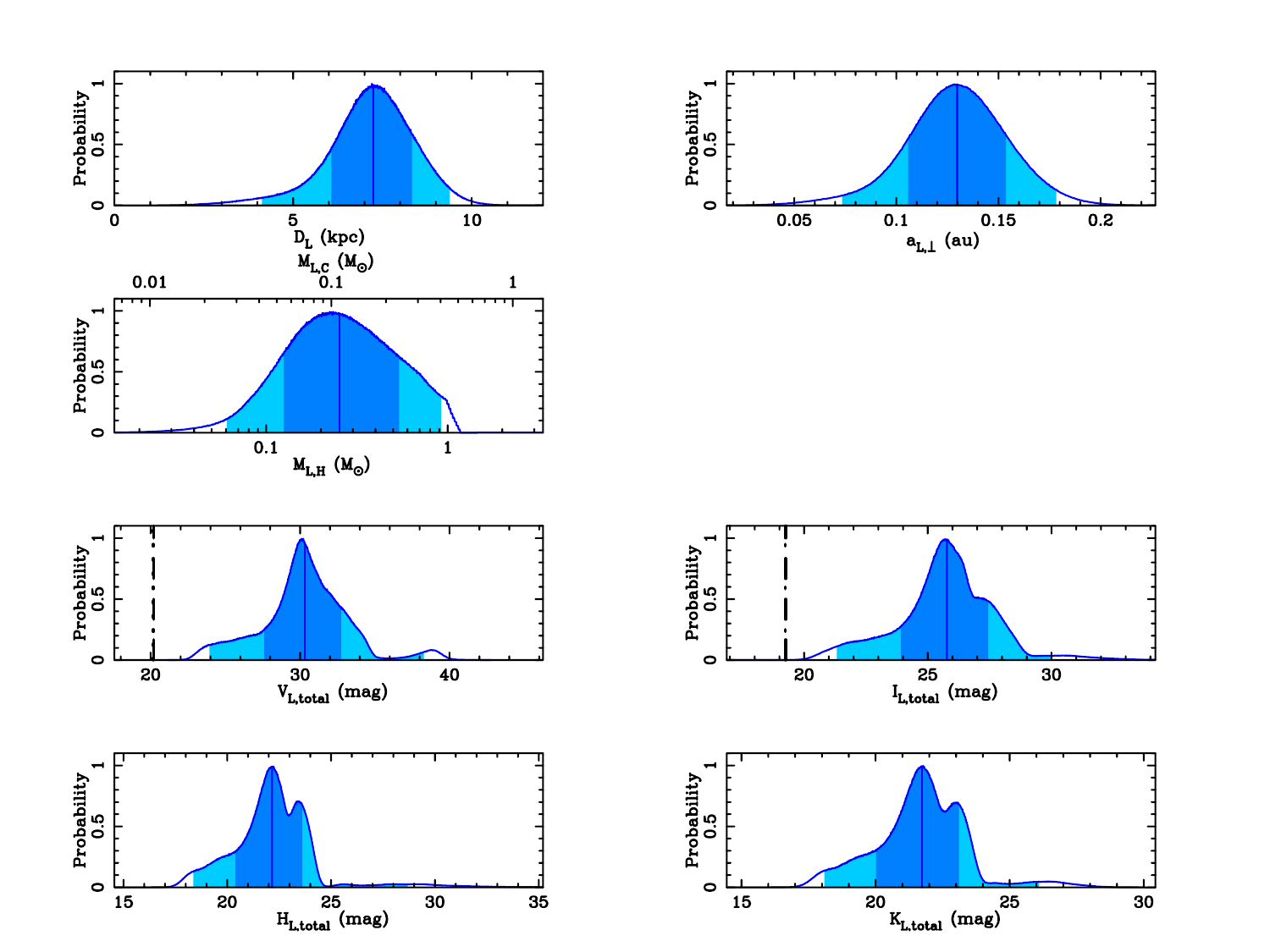}
    \caption{
    Posterior probability distribution of the properties of the lens system by Bayesian analysis for XLclose2.
    In each panel, the dark blue region indicates the $68.3\%$ credible interval, light blue region indicates the $95.4\%$ credible interval, and the blue vertical line indicates the median value.
    The dashed lines at the left end of the panel of apparent $V$- and $I$-band magnitudes with extinction are the blending magnitudes obtained from light curve modeling and are considered as the upper limit of brightness of the lens system.
    }
    \label{fig:Lens_XLclose2}
\end{figure*}

\begin{figure*}
    \centering
    \includegraphics[scale=0.69]{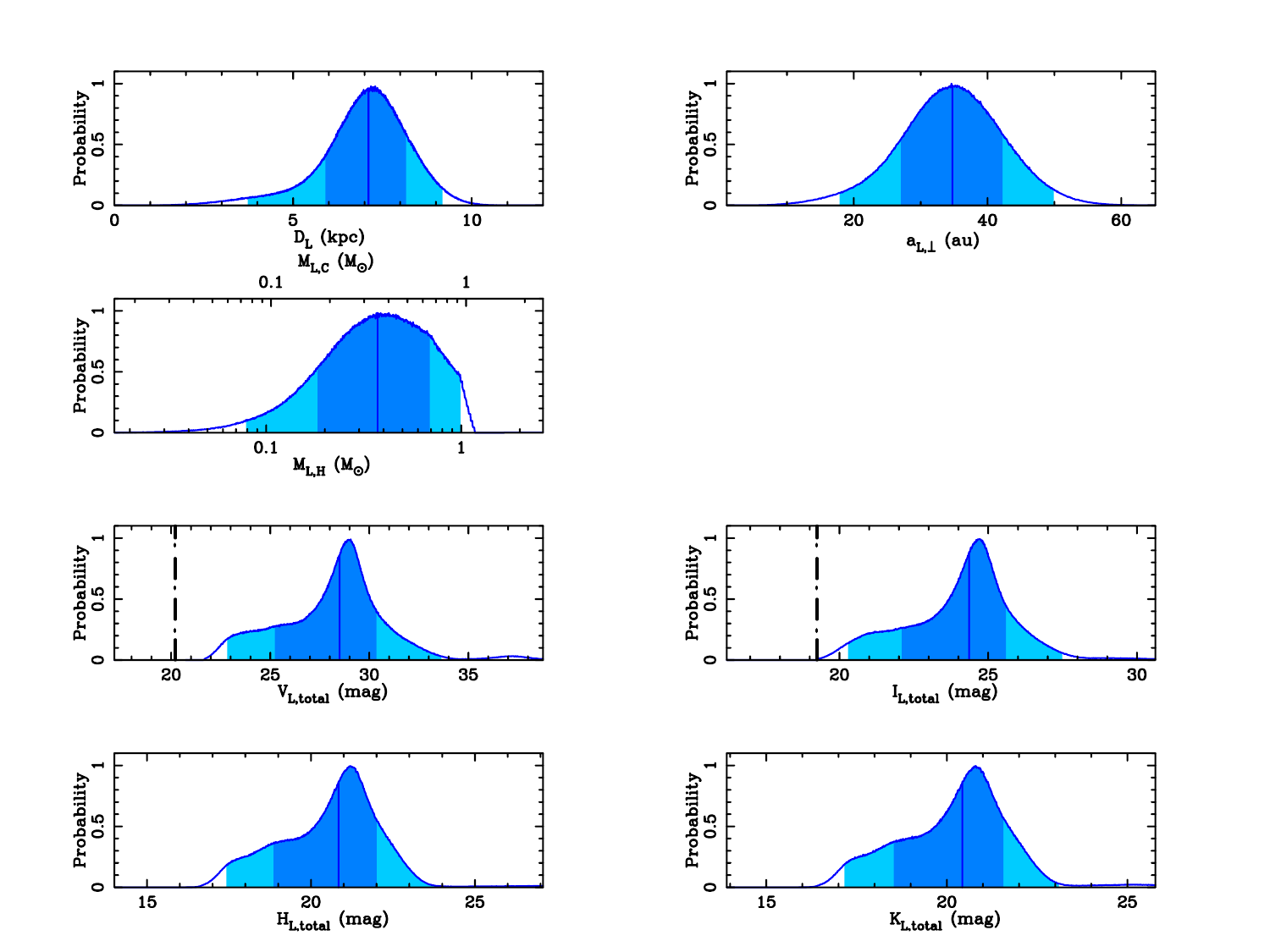}
    \caption{
    Same as Figure~\ref{fig:Lens_XLclose2}, but for XLwide2.
    }
    \label{fig:Lens_XLwide2}
\end{figure*}

\begin{deluxetable*}{c|ccccccccccc}[t!]
\tablecaption{Lens system properties of the 2L1S + xallarap models 
\label{tab:Lens_xallarap}}
\tablewidth{0pt}
\tablehead{Model & XLclose1 & XLclose2 & XLwide1 & XLwide2 \\
\multicolumn{1}{c|}{range of $q$} & \multicolumn{1}{c}{$q\leq0.1$} & \multicolumn{1}{c}{$0.1<q\leq1$} & \multicolumn{1}{c}{$q\leq0.1$} & \multicolumn{1}{c}{$0.1<q\leq1$}
}
\startdata
$D_{\rm L}$ (kpc) & $7.27_{-1.17}^{+1.10}$ & $7.24_{-1.17}^{+1.09}$ & $7.24_{-1.18}^{+1.09}$ & $7.12_{-1.214}^{+1.05}$\\
$M_{\rm L,H}$ ($M_\odot$) & $0.23_{-0.12}^{+0.28}$ & $0.25_{-0.13}^{+0.29}$ & $0.26_{-0.13}^{+0.29}$ & $0.37_{-0.19}^{+0.32}$\\
$M_{\rm L,C}$ ($M_\odot$) & $0.02_{-0.01}^{+0.03}$ & $0.11_{-0.06}^{+0.13}$ & $0.03_{-0.01}^{+0.03}$ & $0.35_{-0.18}^{+0.30}$\\
$a_{\rm {L,\perp}}$ (au) & $0.20_{-0.04}^{+0.04}$ & $0.13_{-0.02}^{+0.02}$ & $11.55_{-2.09}^{+2.03}$ & $34.74_{-7.69}^{+7.51}$\\
$a_{\rm L,exp}$ (au) & $0.25_{-0.06}^{+0.13}$ & $0.16_{-0.04}^{+0.08}$ & $13.91_{-3.15}^{+7.39}$ & $42.10_{-10.92}^{+21.78}$\\
$V_{\rm L,H}$ (mag) & $30.61_{-2.54}^{+2.56}$ & $30.34_{-2.66}^{+2.42}$ & $30.25_{-2.71}^{+2.31}$ & $29.17_{-3.38}^{+1.83}$\\
$I_{\rm L,H}$ (mag) & $26.09_{-1.75}^{+1.66}$ & $25.90_{-1.83}^{+1.57}$ & $25.84_{-1.86}^{+1.51}$ & $25.05_{-2.35}^{+1.22}$\\
$H_{\rm L,H}$ (mag) & $22.53_{-1.73}^{+1.27}$ & $22.35_{-1.80}^{+1.26}$ & $22.30_{-1.82}^{+1.23}$ & $21.53_{-2.02}^{+1.17}$\\
$K_{\rm L,H}$ (mag) & $22.11_{-1.70}^{+1.21}$ & $21.93_{-1.75}^{+1.20}$ & $21.87_{-1.77}^{+1.18}$ & $21.12_{-1.94}^{+1.14}$\\
$V_{\rm L,C}$ (mag) & $41.53_{-1.76}^{+0.98}$ & $33.68_{-2.97}^{+6.20}$ & $41.42_{-2.13}^{+1.03}$ & $29.35_{-3.13}^{+1.88}$\\
$I_{\rm L,C}$ (mag) & $35.24_{-3.05}^{+1.75}$ & $28.13_{-1.94}^{+3.32}$ & $34.85_{-3.13}^{+1.88}$ & $25.18_{-2.16}^{+1.25}$\\
$H_{\rm L,C}$ (mag) & $33.74_{-3.99}^{+3.00}$ & $24.12_{-1.50}^{+5.98}$ & $33.42_{-4.24}^{+3.12}$ & $21.66_{-1.92}^{+1.16}$\\
$K_{\rm L,C}$ (mag) & $30.45_{-2.24}^{+1.47}$ & $23.63_{-1.44}^{+3.68}$ & $30.00_{-2.24}^{+1.54}$ & $21.25_{-1.85}^{+1.14}$\\
$V_{\rm L,total}$ (mag) & $30.60_{-2.54}^{+2.55}$ & $30.29_{-2.68}^{+2.46}$ & $30.25_{-2.71}^{+2.31}$ & $28.50_{-3.27}^{+1.85}$\\
$I_{\rm L,total}$ (mag) & $26.08_{-1.75}^{+1.65}$ & $25.77_{-1.86}^{+1.67}$ & $25.83_{-1.86}^{+1.51}$ & $24.36_{-2.26}^{+1.24}$\\
$H_{\rm L,total}$ (mag) & $22.53_{-1.73}^{+1.27}$ & $22.16_{-1.76}^{+1.44}$ & $22.30_{-1.82}^{+1.22}$ & $20.84_{-1.98}^{+1.16}$\\
$K_{\rm L,total}$ (mag) & $22.10_{-1.70}^{+1.21}$ & $21.73_{-1.71}^{+1.38}$ & $21.87_{-1.77}^{+1.18}$ & $20.43_{-1.89}^{+1.14}$\\
$V_{\rm Blend}$ (mag) & $20.21 \pm  0.03$ & $20.21 \pm 0.03$ & $20.21 \pm 0.03$ & $20.21 \pm 0.03$\\
$I_{\rm Blend}$ (mag) & $19.25 \pm  0.01$ & $19.25 \pm 0.01$ & $19.25 \pm 0.01$ & $19.25 \pm 0.01$\\
\hline \hline
$\chi^2$ & 10856.4 & 10840.9 & 10861.2 & 10842.7\\
$\Delta\chi^2$ & 15.5 & - & 20.3 & 1.8\\
\enddata
\end{deluxetable*}
The distance from the Earth to the lensing system, $D_{\rm L}$, and the total mass of the host and companion in the lensing system, $M_{\rm L}$, can be described by the following equations \citep{Gaudi2012},

\begin{equation}
\label{eq:D_L}
    D_{\rm L}=\frac{\rm au}{\pi_{\rm E}\theta_{\rm E}+\pi_{\rm S}},
\end{equation}

\begin{equation}\label{eq:M_L}
    M_{\rm L}=\frac{\theta_{\rm E}}{\kappa \pi_{\rm E}},
\end{equation}

\noindent
where $\kappa = 4G/(c^2 {\rm au}) \sim 8.144$ [mas  $M_{\odot}^{-1}$] and $\pi_{\rm S}$ is the parallax of the source star written as $\pi_{\rm S}={\rm au}/D_{\rm S}$.

Since the parallax effect was not detected in this event, we conducted a Bayesian analysis \citep{Beaulieu+2006,Gould+2006,Bennett+2008} to estimate the parameters of the lens system for the 2L1S + xallarap models.
For the prior probability distributions, we used the mass density and velocity distributions of the Galaxy model from \citet{Han+1995}, and we used the mass function from \citet{Sumi+2011}.
Since the prior distribution only considers a single star, we scaled the event timescale and the Einstein radius to match those of the lens host so that the physical parameters of the lens host and companion can be properly estimated.
The event timescale of the lens host $t_{\rm {E,H}}$ and the Einstein radius of the lens host $\theta_{\rm {E,H}}$ are expressed using the mass ratio $q$ as follows:

\begin{equation}
\label{eq:t_E,H}
    t_{\rm {E,H}}=\frac{t_{\rm E}}{\sqrt{1+q}},
\end{equation}

\begin{equation}\label{eq:theta_E,H}
    \theta_{\rm {E,H}}=\frac{\theta_{\rm E}}{\sqrt{1+q}}.
\end{equation}

We also estimated the apparent magnitudes of the lens system in the $V$-, $I$-, $K$-, and $H$-bands with extinction.
The magnitudes were obtained using the mass-luminosity relation for main-sequence stars \citep{Henry+1993,Kroupa+1997} and the isochrone model for 5 Gyr old sub-stellar objects \citep{Baraffe+2003}.
The blending flux $f_{\rm b}$ from the light curve modeling was used as the upper limit of the lens brightness.
Following \citet{Bennett+2015}, we estimated the extinction in front of the lens using the following equation:

\begin{equation}\label{eq:extinction}
    A_{i, \rm {L}}=\frac{1-\exp{[-D_{\rm L}/h_{\rm {dust}}}]}{1-\exp{[-D_{\rm S}/h_{\rm {dust}}}]}A_{i, \rm {S}},
\end{equation}

\noindent
where $i$ corresponds to the observed wavelength band, $A_{i, \rm {L}}$ is the total extinction in the $i$-band of the lens, $A_{i, \rm {S}}$ is the total extinction in the $i$-band of the source, $h_{\rm {dust}}$ is the scale length of dust in the event direction, given by $h_{\rm {dust}}=(0.1 \: {\rm {kpc}})/\sin {|b|}$ as a function of the Galactic latitude $b$ of the event.
We estimated $A_{\rm H}$ and $A_{\rm K}$ from $A_{\rm V}$ using the wavelength dependence of extinction law in the direction of the Galactic center from \citet{Nishiyama+2008}.

Table~\ref{tab:Lens_xallarap} lists the estimated parameters: the distance from the Earth to the lens, $D_{\rm L}$; the lens host mass, $M_{\rm {L,H}}$; the lens companion mass, $M_{\rm {L,C}}$; the orbital radius projected to the observation plane, $a_{\rm L, \perp}$; the expected orbital radius, $a_{\rm L, exp}$; the magnitudes with the extinction in the four wavelength bands $V_{{\rm L},j}$, $I_{{\rm L},j}$, $H_{{\rm L},j}$ and $K_{{\rm L},j}$ where $j$ consists of ``H" for the lens host, ``C" for the lens companion and `total" for the host and companion combined; the magnitudes of the blends in the $V$- and $I$-bands which are the upper limits of brightness in the lens system, $V_{\rm {blend}}$, $I_{\rm {blend}}$.
Figures~\ref{fig:Lens_XLclose2} and \ref{fig:Lens_XLwide2} show the posterior probability distributions for XLclose2 and XLwide2, respectively.
The distribution of XLclose2 indicates a M-type or K-type stellar binary with a projected orbital radius $a_{\rm {L,\perp}} = 0.13_{-0.02}^{+0.02}$ au located $7.2_{-1.2}^{+1.1}$ kpc from the Earth.
The distribution of XLwide2 also indicates a M-type or K-type stellar binary with a projected orbital radius $a_{\rm {L,\perp}} = 34.74_{-7.69}^{+7.51}$ au located $7.1_{-1.2}^{+1.0}$ kpc from the Earth. 
Comparing the properties of the lens systems of the four models listed in Table~\ref{tab:Lens_xallarap}, while the parameters related to the companion differ significantly among the models, they are consistent in that the stellar type and the distance from the Earth.

As described in Section~\ref{sec:Source_Lens_Properties}, the apparent magnitude of the source for XLclose2 is $(H_{\rm S}, K_{\rm S})=(18.52\pm0.49, 18.29\pm0.48)$.
The apparent magnitude for the lens host and lens companion combined is $(H_{\rm {L,total}}, K_{\rm {L,total}})=(22.16\pm1.16, 21.73\pm1.55)$. 
Therefore, XLclose2 has a contrast between the apparent lens brightness and the apparent source brightness where $3.6\pm1.7$ mag in the $H$-band and $3.4\pm1.6$ mag in the $K$-band.
The XLclose1 and XLwode1 models also have similar contrast to XLclose2, respectively.
On the other hand, the contrast between the apparent lens brightness and the apparent source brightness in the XLwide2 model is $2.3\pm1.6$ in the $H$-band and $2.1\pm1.6$ in the $K$-band, slightly lower contrast than that in XLclose2.

\section{Discussion and Conclusion}
\label{sec:Discussion_and_Conclusion}
We performed a detailed analysis of the planetary microlensing candidate, OGLE-2019-BLG-0825.
We first found that there are  systematic residuals with the best fit standard binary model with planetary mass ratio $q \sim 10^{-3}$.  
Therefore, we examined various combinations of possible higher-order effects.
As a result, we found that models which include the xallarap effect can fit the residuals significantly better than models which do not.

Our Bayesian analysis for the best model XLclose2 estimated the lens host mass to be $0.25_{-0.13}^{+0.29}$ $M_\odot$ and the lens system to be located $7.24_{-1.17}^{+1.09}$ kpc from Earth.
For XLwide2, which is the best solution at $s>1$, the lensing host is $0.37_{-0.19}^{+0.32}$ $M_\odot$, and the lens system is estimated to be located $7.12_{-1.22}^{+1.05}$ kpc from Earth.
Owing to degenerate solutions with various combinations of $(q,s)$ values, the uncertainties in the mass and orbital radius of the lens companion are large.
Since the relative proper motion between the lens and the source is about 1 mas $\rm { {yr}}^{-1}$ and the apparent magnitude contrast is large, it will be more than 30 years before the source and lens can be observed separately with the current high-resolution imaging instruments.
In adaptive optics (AO) observations by The European Extremely Large Telescope (ELT), the FWHM is expected to reach 10 mas in the $H$-band and 14 mas in the $K$-band \citep{Ryu+2022}.
Therefore, it may be possible to observe the source and lens separately by mid-2030.
It is unlikely that the degeneracy of the models will be resolved by follow-up observations because the proper motion and brightness of the lens system are comparable across models, but it may constrain the uncertainty in the lens system properties somewhat.

Calculations applying the $D_{\rm S}=8.0 \pm 1.4$ kpc assumption and the isochrone model with age 10 Gyr in solar metallicity to the source show that the source companion OGLE-2019-BLG-0825Sb in the best 2L1S + xallarap model has a semi major axis of $0.0594 \pm 0.0005$ au and an orbital period of $5.53 \pm 0.05$ days with mass $0.048 \pm 0.004 $ $M_\odot$ orbiting the host source star OGLE-2019-BLG-0825S.
The mass of the source companion is about that of a brown dwarf.
The $I$-band luminosity ratio of the companion to the host is $L_{\rm S,C}/L_{\rm S,H} = (1.0\pm0.3) \times {10}^{-7}$, which is faint and consistent with this analysis where the magnified flux of the second source is too weak to be detected.
We note that these properties of the source system are almost the same among the various models considered, even though the parameters of the lens system change.

We considered whether a variable source star could also explain the $\sim5$ day luminosity variations of this event, without using the xallarap effect.
Most of Classical Cepheids have a pulsation periods ranging from about 1 to 100 days, and the longest period ones being rare, with a pulsation amplitude in $I$-band of $0.05-1$ mag \citep{Klagyivik+2009}, and the following period-luminosity relations \citep{Gaia+2017}:

\begin{equation}\label{eq:PL_claccical_cepheid}
    M_I=-2.98 \log P – (1.28 \pm 0.08); \sigma_{\rm {rms}}=0.78,
\end{equation}

\noindent
where $\sigma_{\rm {rms}}$ is the variance around the periodic luminosity relation.
At a pulsation period $P=5.50 \pm0.05$ days, the absolute magnitude of a type I Cepheid would be $M_I =-3.48 \pm 0.08$ mag.
However our estimated absolute magnitude is $M_I = 4.5 \pm 0.4$ mag, which is too faint for a classical Cepheid (see Table~\ref{tab:Source_xallarap}).
Type $\mathrm{I}\hspace{-1.2pt}\mathrm{I}$ Cepheids have a pulsation period of about 1 to 50 days, with a pulsation amplitude of $0.3-1.2$ mag, and the following period-luminosity relationships \citep{Ngeow+2022}:

\begin{equation}\label{eq:PL_claccical_type2}
    M_I=-(2.09 \pm 0.08) \log P – (0.39 \pm 0.08); \sigma_{\rm {rms}}=0.24.
\end{equation}

\noindent
For a pulsation period $P=5.50 \pm 0.05$ days, the absolute magnitude of a type $\mathrm{I}\hspace{-1.2pt}\mathrm{I}$ Cepheid would be $M_I=-1.94 \pm 0.13$ mag, which is also not plausible.
RR Lyrae variables have color magnitudes close to main-sequence stars, but with a pulsation period of less than one day \citep[e.g.,][]{Soszynski+2009}.
Delta Scuti variables have a pulsation period of $0.01-0.2$ days, and Gamma Doradus variables have a pulsation period of 0.3-2.6 days, both shorter than the xallarap signal of 5 days, and the spectral type is $A-F$, which is blue compared to the color of the source of this event.
Furthermore, as described in Section \ref{subsec:Xallarap}, we performed a fitting with a model with variable source flux, using the best standard 2L1S model (i.e., close1).
However, the improvement from the best standard 2L1S model was only $139.1$, $\Delta \chi^2 = 764.6$ worse than the 2L1S + xallarap model.
Therefore, we conclude that it is difficult to explain the xallarap signal assuming a variable source star.
Note that although the conclusion is that the source of this event is not a variable star, many variable stars in the direction of the Galactic bulge have been discovered \citep[e.g.,][]{Soszynski+2011,Iwanek+2019}, and there is a possibility that a candidate planetary microlensing event with a variable source will be observed in the future.

For the lens system, the inclusion of the xallarap effect significantly changed the $\Delta \chi^2$-plane of the mass ratio $q$ vs. separation $s$.
The mass ratio of the best model was $q = (3.3 \pm 0.1) \times 10^{-3}$ without accounting for a xallarap effect, but became $q = (4.4 \pm 1.1) \times 10^{-1}$ with the xallarap effect.
Furthermore, degenerate solutions with various combination of $(q,s)$ values were found within a small range of $\Delta\chi^2 \lesssim 10$.
This event is the first case that the short-period xallarap effect significantly affects the binary-lens parameters $q,s$.
This effect is most likely to be seen in events with a caustic or cusp approach and no clear sharp caustic crossing.
In events with a clear sharp caustic crossing, this effect is not significant because the mass ratio $q$ and separation $s$ can be constrained from the caustic shape.

Although the xallarap effect has been examined in the past \citep[e.g.,][]{Bennett+2008, Sumi+2010}, few events have been able to eliminate possibilities of systematic errors and clearly identify the xallarap signal.
\citet{Miyazaki+2020} analyzed a planetary microlensing event OGLE-2013-BLG-0911 and found a significant xallarap signal. 
They conclude from the fitting parameters that the source companion, OGLE-2013-BLG-0911Sb has a mass $M_{\rm S,C} = 0.14 \pm 0.02$ $M_\odot$, an orbital period $P_\xi = 36.7 \pm 0.8$ days and a semi-major axis $a_{\rm S} = 0.225 \pm 0.004$ au.
However, they assume $M_{\rm S,H} = 1$ $M_\odot$ and $D_{\rm S}=8$ kpc.
Recently \citet{Rota+2021} analyzed a candidate planetary event MOA-2006-BLG-074 and detected a xallarap effect.
They estimated the source host's mass $M_{\rm S,H} = 1.32 \pm 0.36$ $M_\odot$ from the color and magnitude of the source and found that the companion with mass $M_{\rm S,C} = 0.44 \pm 0.14$ $M_\odot$ is orbiting the source host with orbital period $P_\xi = 14.2 \pm 0.2$ days and semi-major axis $a_{\rm S} = 0.043 \pm 0.012$ au.
The OGLE-2019-BLG-0825 event in this work is the second case after the MOA-2006-BLG-074 event \citep{Rota+2021}, in which the physical properties of a source system were estimated from the color and magnitude of the source.
This event will be a valuable example for future xallarap microlensing analyses.

\citet{Rahvar+2009} suggested that planets orbiting sources in the Galactic bulge could be detected by the xallarap effect with sufficiently good photometry.

The fraction of close binaries like OGLE-2013-BLG-0911Sb is known to be anticorrelated with metallicity \citep{Moe+2019}.
The Galactic bulge observed in microlensing surveys suggests the presence of super-solar, solar, and low-metallicity components with [Fe/H]$\sim 0.32$, [Fe/H] $ \sim 0.00$, and [Fe/H] $ \sim -0.46$, respectively \citep{Garcia+2018}. 
\citet{Moe+2019} reported that the fraction of close binaries, $F_{\rm close}$, with separation $a<10$ au is $F_{\rm close} = 24\% \pm 4 \%$ at [Fe/H]$=-0.2$ and $F_{\rm close} = 10\% \pm 3 \%$ at [Fe/H]$=0.5$.
However, the occurance ratio of a companion with an orbit even shorter than $\sim$0.5 au, to which the xallarap effect has sensitivity, is poorly understood.

\citet{Tokovinin+2006} found that $\sim68\%$ of close binary systems in the solar neighborhood with orbital period $P=3-6$ days have an outer tertiary companion.
\citet{Eggleton+2006} and \citet{Fabrycky+2007} showed that Kozai-Lidov cycles with tidal friction \citep[KCTF;][]{Kiseleva+1998,Eggleton+2001} produce such very close binaries.
First, in the KCTF, the inner companion's eccentricity is pumped by perturbations from the outer tertiaries.
The inner companion in the eccentric orbit undergoes tidal friction near the periastron, and the orbit of the inner companion is finally circularized.
Timescale equations for tidal circularization have been studied \citep[e.g.,][]{Adams+2006,Correia+2020}.
Because of their small radius relative to their mass the orbits of brown dwarfs are expected to take longer to circularize than those for Jupiter-like planets with the same orbital period over the Gyr scale.
However this is difficult to estimate because the tidal quality factor for brown dwarfs is not well constrained \citep{Heller+2010,Beatty+2018}.
Meanwhile, \citet{Meibom+2005} demonstrated from the distribution of orbital eccentricity vs. orbital period that most of the companions are circularized when the orbital period is shorter than $\sim$15 days for the companions of halo stars and $\sim$10 days for the companions of nearby G-type primaries.
Therefore, in this analysis of OGLE-2019-BLG-0825, the source orbital eccentricity was fixed to $e_{\xi}=0$.
We also performed an analysis with free eccentricity, but our results were almost the same, and the improvement in $\chi^2$ was only $\Delta\chi^2\sim16$, despite two additional parameters, $e_{\xi}$ and $T_{\rm peri}$.

Disc fragmentation and migration are also possible formation processes for close binaries.
\citet{Moe+2018} noted that the close binary fraction of solar-mass, pre-main-sequence binaries and field main-sequence binaries is almost identical \citep{Mathieu+1994,Melo+2003}, and concluded that majority of very close binaries with semi-major axis $a$ $<$ 0.1 au migrated when there was still gas composition in the circumstellar disc.
Furthermore, \citet{Moe+2019} showed that $90\%$ of close binary stars with $a$ $<$ 10 au are the product of disc fragmentation.
\citet{Tokovinin+2020} use simulations of disc fragmentation to show that the companion has difficulty migrating to $P$ $<$ 100 days without undergoing accretion that would grow it to more than 0.08 $M_\odot$, explaining brown dwarf deserts.

The source companion OGLE-2019-BLG-0825Sb is the least massive source companion found in a xallarap event, and our favored interpretation is that it has a brown dwarf mass.
The occurrence rate for brown dwarfs orbiting main-sequence stars have been found to be low, less than $1\%$ \citep{Marcy+2000,Grether+2006,Sahlmann+2011,Santerne+2016,Grieves+2017}.
Fewer than 100 brown dwarf companions have been found in solar-type stars \citep[e.g.,][]{Ma+2014,Grieves+2017}. 
There is a particularly dry region at orbital period $P$ $<$ 100 days \citep[e.g.,][]{Kiefer+2019,Kiefer+2021}. 
Therefore, if OGLE-2019-BLG-0825Sb is a short-period brown dwarf, it is a resident of the driest region of the brown dwarf desert, making it a very valuable sample for studying brown dwarf formation.
\citet{Miyazaki+2021} estimated the planetary yield detected by the Nancy Grace Roman Space Telescope \citep[][previously named WFIRST, hereafter Roman]{Spergel+2015} via xallarap signals assuming a planetary distribution of masses and orbital periods of \citet{Cumming+2008}.
They predicted that Roman will characterize tens of short-period Jupiter - brown dwarf mass companions such as OGLE-2019-BLG-0825S.
By comparing the predictions with the actual results, it will be possible to verify the brown dwarf desert in the Galactic bulge.

%
In this study, we assumed $D_{\rm S}=8.0$ $\pm 1.4$ kpc.
Roman observations may be able to measure $D_{\rm S}$ by directly measuring astrometric parallax for bright source events \citep{Gould+2015}.
Even for non-bright source events, $D_{\rm S}$ can be determined by measuring the lensing flux $F_{\rm L}$, $\pi_{\rm E}$, and $\theta_{\rm E}$. 
Events with photometric accuracy $\leq 0.01$ mag have been analytically shown to have the potential to measure $\theta_{\rm E}$ with $\leq 10 \%$ accuracy via astrometric microlensing observations in space \citep{Gould+2014}.
Future observations of the xallarap effect may reveal the distribution of short-period binary stars in the Galactic center, which are usually difficult to observe.
\newline
\newline
Y.K.S. acknowledges the financial support from the Hayakawa Satio Fund awarded by the Astronomical Society of Japan.
Work by N.K. is supported by the JSPS overseas research fellowship.
The MOA project is supported by JSPS KAKENHI Grant Number JSPS24253004, JSPS26247023, JSPS23340064, JSPS15H00781, JP16H06287, and JP17H02871.
This research has made use of the KMTNet system operated by the Korea Astronomy and Space Science Institute (KASI) and the data were obtained at three host sites of CTIO in Chile, SAAO in South Africa, and SSO in Australia.
Y.S. acknowledges support from BSF Grant No. 2020740.
J.C.Y. acknowledges support from U.S. NSF Grant No. AST-2108414.
U.G.J., N.B-M. and J.S. acknowledge funding from the European Union H2020-MSCA-ITN-2019 under Grant Number 860470 (CHAMELEON), from the Novo Nordisk Foundation Interdisciplinary Synergy Program Grant Number NNF19OC0057374 and from the Carlsberg Foundation under Grant Number CF18-0552. 
N.P.'s work was supported by Funda\c{c}\~ao para a Ci\^{e}ncia e a Tecnologia (FCT) through the research grants UIDB/04434/2020 and UIDP/04434/2020.
P.L.P. was partly funded by ``Programa de Iniciación en Investigación-Universidad de Antofagasta INI-17-03".
G.D. acknowledges support by ANID, BASAL, FB210003. 
\bibliography{reference}{}
\bibliographystyle{aasjournal}
\end{document}